%% file: tlpff.tex
\documentclass[epj]{svjour}
\date{March 2010}
\usepackage{epsfig}
\usepackage{latexsym}
\usepackage{amssymb}
\usepackage{booktabs}
\usepackage{graphicx}
\usepackage{bm}
\usepackage{david}

\newcommand{\be}{\begin{equation}}
\newcommand{\ee}{\end{equation}}
\newcommand{\ba}{\begin{eqnarray}}
\newcommand{\ea}{\end{eqnarray}}
\newcommand{\bi}{\begin{itemize}}
\newcommand{\ei}{\end{itemize}}
\newcommand{\ud}{\,\mathrm{d}}

\newcommand{\re}{\mathop{\rm Re}}

\newcommand{\nn}{\nonumber \\}

\newcommand{\half}{{\textstyle\frac{1}{2}}}

\newcommand{\<}{\langle}
\renewcommand{\>}{\rangle}
\newcommand{\eq}{Eq.~}

\newcommand{\fig}{Fig.~}
\newcommand{\tab}{Tab.~}
\newcommand{\la}{\label}

\newcommand{\txts}{\textstyle}

\newcommand{\bx}{\boldsymbol{x}}

\newcommand{\bq}{\boldsymbol{q}}
\newcommand{\bk}{\boldsymbol{k}}

\newcommand{\Pihat}{\widehat\Pi}
\newcommand{\amuHLO}{a^{\rm HLO}_\mu}

\begin{document}
\title{Vector Correlators in Lattice QCD: methods and applications}

\author{David Bernecker\inst{1} \and Harvey~B.~Meyer\inst{1}}
\institute{Institut f\"ur Kernphysik,
 Johannes Gutenberg Universit\"at Mainz, 
 55099 Mainz, Germany}

\date{\today}

\abstract{ 
We discuss the calculation of the leading hadronic vacuum polarization
in lattice QCD. Exploiting the excellent quality of the compiled
experimental data for the $e^+e^-\to{\rm hadrons}$ cross-section, we
predict the outcome of large-volume lattice calculations at the
physical pion mass, and design computational strategies for the
lattice to have an impact on important phenomenological quantities
such as the leading hadronic contribution to $(g-2)_\mu$ and the
running of the electromagnetic coupling constant.  First, the $R(s)$
ratio can be calculated directly on the lattice in the threshold
region, and we provide the formulae to do so with twisted boundary
conditions.  Second, the current correlator projected onto zero
spatial momentum, in a Euclidean time interval where it can be
calculated accurately, provides a potentially critical test of the
experimental $R(s)$ ratio in the region that is most relevant for
$(g-2)_\mu$.  This observation can also be turned around: the vector
correlator at intermediate distances can be used to determine the
lattice spacing in fm, and we make a concrete proposal in this
direction. Finally, we quantify the finite-size effects on the current
correlator coming from low-energy two-pion states and provide a
general parametrization of the vacuum polarization on the torus.
}

\maketitle


\input{intro}
\input{compar}

\input{finivol}

\input{piQ2}

\input{tau0}

\input{twisted}
\input{concl}

\subsection*{Acknowledgments}
We thank Hartmut Wittig, Michele Della Morte, Benjamin J\"ager and 
Andreas J\"uttner for helpful discussions and for providing the lattice data
displayed in \fig(\ref{fig:PiCompar}). HBM also thanks Achim Denig for 
helpful discussions. HBM's work is supported by the 
\emph{Center for Computational Sciences} in Mainz.

\appendix

\input{appendixFSE}
\input{potwell}

\bibliographystyle{JHEP}
\bibliography{/home/meyerh/CTPHOPPER/ctphopper-home/BIBLIO/viscobib,../../DAVIDBERNECKER/zusatz}

\end{document}

%% file: intro.tex
\section{Introduction}

In quantum field theory, the information encoded in the correlation
functions of conserved currents has important phenomenological
applications.  The correlation function of the electromagnetic
current, in particular, quantifies the polarization of the `vacuum' by
virtual particles induced by the passage of a photon. This
virtuality-dependent vacuum polarization $\Pi(Q^2)$ affects the
propagation of the photon and has physically observable consequences,
see for instance~\cite{Jegerlehner:2001wq}. One of them is the running
of the fine structure constant $\alpha(Q^2)$, which now depends on the
four-momentum squared of the photon via \eq(\ref{eq:alpha}) below.
Another consequence of the polarization of the vacuum is a
contribution of all virtual particles to the magnetic moments of
leptons, which can be measured to very high precision in the case of
the electron and the muon (see~\cite{Jegerlehner:2009ry} for a review
of the subject). We will discuss this application extensively below.

One important contribution to the vacuum polarization comes from QCD.
At high virtuality $Q^2$, this contribution to $\Pi(Q^2)$ is
calculable in perturbation theory, due to the asymptotic freedom
property of QCD. Below a scale of a few GeV, the vacuum polarization
receives large non-perturbative contributions, making it inaccessible
to known analytic methods. From here on, we will focus exclusively on
the QCD contribution to $\Pi(Q^2)$, but will keep using the same
symbol.  In particle phenomenology, it has been customary to extract
the low-$Q^2$ part of the function $\Pi(Q^2)$ from experiments via a
dispersion relation (\eq\ref{eq:DispRelR} below). However it is also
possible to directly calculate $\Pi(Q^2)$ from first principles using
numerical lattice QCD
methods~\cite{Blum:2002ii,Aubin:2006xv,Feng:2011zk,DellaMorte:2010sw},
roughly for a range of momenta $0.1{\rm GeV}^2\lesssim Q^2\lesssim 4.0{\rm
  GeV}^2$.  What limits the upper end of the momentum range is the
size of the lattice spacing $a$, whose inverse provides a momentum
cutoff $\sim\pi/a$.  On the lower end, it is the discreteness of the
available momenta in a finite volume ($|Q_{\rm min}|=\frac{2\pi}{L}$
on a torus of dimension $L$) that limits the accessible $Q^2$ values.
We note that current correlators involving heavy flavors of quarks
have been used on the lattice for other purposes, namely determining
the charm quark mass and the QCD coupling
constant~\cite{Allison:2008xk}.

To obtain the leading hadronic contribution to the anomalous magnetic
moment of the muon $\amuHLO$, the imaginary part of the vacuum
polarization at timelike momenta is folded with an analytically known
QED kernel~\cite{Lautrup:1971jf} which involves only the scale
$m_\mu$. The same quantity $\amuHLO$ can also be expressed as an
integral over spacelike momenta~\cite{Lautrup:1969nc}.  
An important observation of Blum~\cite{Blum:2002ii} was that this opens 
the possibility to evaluate it in the Euclideanized theory, which can 
be simulated non-perturbatively by Monte-Carlo
methods~\cite{Blum:2002ii,Aubin:2006xv,Feng:2011zk,DellaMorte:2010sw}.
Because the muon mass $\mu\approx 105$MeV is small on hadronic scales,
the contribution to $a_\mu$ is dominated by the region of small $Q^2$
or, alternatively, by long distance contributions.  How exactly this
affects the calculation is one topic we will address.

Since the vector spectral function is
already extremely well known from particle physics experiments, only
highly accurate lattice predictions will have a useful impact on
phenomenology.  By the same token, the knowledge of the vector
spectral function allows us to design useful observables which are
both accurately computable on the lattice and critically challenge the
experimental measurements that are relevant to the determination of
$\Delta\alpha(M_Z^2)$ and $\amuHLO$. 

One of our goals will be to understand some of the systematic errors
that may affect the direct lattice calculation of the vacuum
polarization. A second goal is to propose other promising
computational strategies, in increasing level of `ambition', which we
believe could have a phenomenological impact. These strategies
represent a compromise between computational feasability and
phenomenological relevance. The third topic is our proposal of a new
reference scale $\tau_0$ which is both accurately calculable on the
lattice and extractable from experimental data with negligible model
dependence. The time-scale $\tau_0$ is defined from the isospin vector
current correlator.

In section (\ref{sec:prelim}), after a review of the basic relations
between the relevant observables, we provide a comparison of the
vacuum polarization calculated on the lattice and the vacuum
polarization obtained from the phenomenological $R$ ratio by employing
the dispersion relation (\ref{eq:DispRel}). We observe that even
state-of-the-art lattice calculations still yield a vacuum
polarization about a factor of two smaller than phenomenology
indicates in a wide range of momenta.  In section (\ref{sec:finivol}),
we provide the general parametrization of the polarization tensor
$\Pi_{\mu\nu}(Q)$ compatible with the symmetries of the torus and the
conservation of the electromagnetic current; this allows one to
perform several tests of finite-volume effects.  In section
(\ref{sec:mix}) we will switch to the mixed representation $(t,\bq)$,
where $t$ is Euclidean time, which facilitates the interpretation of
the vector correlator in terms of physical states. At large times $t$,
the current correlator is exponentially dominated by the low-lying
states. These are two-pion states on the torus and due to the sparsity
of momenta available to them, their contribution is affected by O(1)
finite volume effects.  Based on this understanding, we will consider
three ways in which the lattice could have an impact on the
determination of $\Delta\alpha(Q^2=M_Z^2)$ and $\amuHLO$. Section
(\ref{sec:tau0}) describes the new reference scale $\tau_0$. In
section (\ref{sec:luescher3d}), we present a technical generalization
of the L\"uscher formula in the vector channel for twisted boundary
conditions (see also~\cite{deDivitiis:2004rf,Kim:2010sd} in the scalar channel), 
which are expected to help in the determination of the timelike pion form
factor. The concluding section contains a summary of our findings and
proposed computational strategies.

\section{Preliminaries and status of 
vacuum polarization determinations\la{sec:prelim}}

We begin this section by introducing the relevant quantities and
reviewing the most important relations among them. This will allow us
to perform a comparison between lattice experimental data.  Our
Minkowski metric convention is ($+---$).  Euclidean momenta are
denoted by a capital letter, Minkowski momenta by a small letter.

The electromagnetic current 
\be
j^{\rm em}_\mu={\txts\frac{2}{3}}\bar u\gamma_\mu u - {\txts\frac{1}{3}} \bar d \gamma_\mu d 
-{\txts\frac{1}{3}} \bar s \gamma_\mu s + \dots
\ee
is the central operator of interest in this paper.
The corresponding spectral function is defined as 
\be \la{eq:rhoDef}
\rho_{\mu\nu}(k)\equiv \frac{1}{2\pi}
\int d^4x\, e^{ik\cdot x}
\<0| [j^{\rm em}_\mu(x), j^{\rm em}_\nu(0)]|0\>.
\ee
Due to current conservation and Lorentz invariance, the tensor structure
of $\rho_{\mu\nu}$ is 
\be \la{eq:rhomunu}
\rho_{\mu\nu}(k) = (k_\mu k_\nu - g_{\mu\nu}k^2) \cdot \rho(k^2).
\ee
The spectral density $\rho$ is non-negative. 
In the free theory for massless quarks of charges $Q_f$, 
it is given by a step function,
\be
\rho(s) = \frac{N_c({\txts\sum_f} Q_f^2)}{12\pi^2}\;  \theta(s)
\qquad \textrm{(free massless quarks)}.
\ee
More generally, $\rho(s)$ is related to experimental observables
by the optical theorem,
\be \la{eq:rhoR}
\rho(s) =\frac{R(s)}{12\pi^2},
\qquad
R(s) \equiv  \frac{\sigma(e^+e^-\to {\rm hadrons})}
 {4\pi \alpha(s)^2 / (3s) } .
\ee
The denominator is the treelevel cross-section $e^+e^-\to\mu^+\mu^-$
in the limit $s\gg m_\mu^2$, and we have neglected QED corrections.
At low energies, the spectral density is given by the 
pion form factor~\cite{Jegerlehner:2009ry},
\be
\rho(s) = \frac{1}{48\pi^2} 
\Big(1-\frac{4m_\pi^2}{s} \Big)^{\frac{3}{2}}  |F_\pi(\sqrt{s})|^2,
\qquad |F_\pi(0)|=1.
\la{eq:RFpi}
\ee
This relation holds near threshold, $2m_\pi\leq\sqrt{s}\leq 3m_\pi$, and even up to
$4m_\pi$ if the electromagnetic current is replaced by the isospin
current in the definition of $\rho(s)$.
In~\cite{Meyer:2011um}, 
a formula relating the pion form factor to a finite-volume matrix 
element calculable in lattice QCD was derived,
\be
|F_\pi(E)|^2 = \Big(q \phi'(q) + k \frac{\partial\delta_1(k)}{\partial k}\Big)
\frac{3\pi E^2}{2k^5}    |A_\psi|^2 .
\la{eq:result1}
\ee
Here $E$ equals the invariant mass of the two pions, $k$ is related to $E$
via  $ E = 2\sqrt{m_\pi^2 + k^2}$, $\delta_{1}$ is the
scattering phase shift in the $p$-wave, isospin $I=1$ channel and $A_\psi$ is a
vector-current matrix element between the vacuum and a unit-norm two-pion state
$|\psi^a_\sigma\>$ of energy $E$ on the torus. 
Finally, $q\equiv \frac{kL}{2\pi}$ and $\phi$ is a
known kinematic function~\cite{Luscher:1991cf}.  The scattering phase
$\delta_{1}(k)$ can be extracted (see~\cite{Feng:2010es,Aoki:2007rd} 
and Refs. therein) from the finite-volume spectrum using
the L\"uscher formula~\cite{Luscher:1990ux,Luscher:1991cf}.

In Euclidean space, the natural object is the polarization tensor
\be
\Pi_{\mu\nu}(Q) \equiv \int d^4x \, e^{iQ\cdot x} \<j_\mu(x) j_\nu(0)\>,
\ee
and O(4) invariance and current conservation imply the tensor structure
\be\la{eq:PimunuQ}
\Pi_{\mu\nu}(Q) = \big(Q_\mu Q_\nu -\delta_{\mu\nu}Q^2\big) \Pi(Q^2).
\ee
The function $\Pi(Q^2)$ can be calculated in lattice
QCD~\cite{Blum:2002ii,Aubin:2006xv,Feng:2011zk,DellaMorte:2010sw}.  The leading
hadronic contribution to the vacuum polarization $e^2\Pi(Q^2)$ in the
spacelike domain can be expressed through the spectral function via a
once-subtracted dispersion relation,
\be \la{eq:DispRel}
 \Pi(Q^2)-\Pi(0) = Q^2 \int_0^\infty \ud s \frac{\rho(s)}{s(s+Q^2)}.
\ee
An important physical application of the vacuum polarization is the 
running of the electromagnetic coupling,
\ba\la{eq:alpha}
\alpha(Q^2) &=& \frac{\alpha}{1-\Delta\alpha(Q^2)},\qquad 
\alpha \equiv \alpha(0),
\\
\Delta \alpha(Q^2) &=& 4\pi\alpha \re\big[ \Pi(Q^2) - \Pi(0) \big].
\nonumber
\ea
In particular the value of the coupling at the scale $Q^2=M_Z^2$ is a
precision observable that, combined with the Fermi constant, the $Z$
boson mass, the quark masses and the Higgs mass, lead to a prediction
for the Weinberg angle, which confronted with its direct measurement
leads to an upper bound on the Standard Model Higgs boson mass (see
for instance~\cite{Jegerlehner:2001wq}).

The O($\alpha^2$) hadronic contribution to the muon anomalous magnetic
moment can be expressed in terms of $\Pi(Q^2)$ as 
\ba \la{eq:amublum2} 
\amuHLO &=& \left(\frac{\alpha}{\pi}\right)^2 
\int_0^\infty {\ud Q^2} K_E(Q^2) \Pihat(Q^2),
\\
\la{eq:Pihat}
 \Pihat(Q^2) &=& 4\pi^2\big[\Pi(Q^2) - \Pi(0)\big],
\ea
with the kernel given by\footnote{Our function $K_E$ matches the function 
$f$ introduced in~\cite{Blum:2002ii}.}~\cite{Blum:2002ii}
\ba
  \label{eq:kerK}
K_E(s) &=& \frac{1}{m_\mu^2}\cdot \hat s\cdot Z(\hat s)^3\cdot 
\frac{1 - \hat s Z(\hat s)}{1 + \hat s Z(\hat s)^2}\,,
\\
Z(\hat s) &=& - \frac{\hat s - \sqrt{\hat s^2 + 4 \hat s}}{2  \hat s},
\quad \hat s = \frac{s}{m_\mu^2}\,.
\ea
\eq(\ref{eq:amublum2}) is used to obtain $\amuHLO$ based on the vacuum
polarization computed on the lattice.


%% file: compar.tex
\subsection{Confronting the vacuum polarization from the lattice and the $R(s)$ ratio via the dispersion relation}
\la{sec:compar}

\begin{table}
\centering
\begin{tabular}{c@{$~~~$}c@{$~~~$}c@{$~~~$}c}
\hline\hline
    &    $C_i$   &  $M_i/{\rm GeV}$   & $\Gamma_i/{\rm GeV}$   \\
\hline
0   &   655.5    & 0.7819    &  0.0358  \\
1   &   8.5      & 0.7650    &  0.130  \\
2   &   11.5     & 0.7820    &  0.00829 \\
3   &   50.0     & 1.0195   &  0.00426  \\
\hline\hline
\end{tabular}
\caption{Parameters used in the parametrization (\ref{eq:Rs}) of the $R(s)$ ratio.}
\la{tab:param}
\end{table}

The vacuum polarization can be easily calculated from the $R(s)$ 
ratio by using the optical theorem (\ref{eq:rhoR}) and the dispersion relation (\ref{eq:DispRel}).
In the Euclidean domain, the relation reads
\begin{equation} \la{eq:DispRelR}
\Pihat(Q^2)=\frac{Q^2}{3}\int_0^\infty\! ds\frac{R(s)}{s(s+Q^2)}.
\end{equation}
By taking the phenomenological determination of $R(s)$, the integral can be evaluated 
and a comparison to the vacuum polarization calculated on the lattice is possible.
For this purpose we parametrize the $R(s)$ ratio using Breit-Wigner curves of the form 
\begin{equation}
  \label{eq:BW}
 f(\sqrt{s})=\frac{C\,\Gamma^2}{4(\sqrt{s}-M)^2+\Gamma^2}
\end{equation}
where the parameters $C,M,\Gamma$ are used to match the height of the resonance 
to the experimental data compiled by the Particle Data Group (PDG) \cite{PDG2008}.
Altogether our parametrization reads, for $s$ in units of GeV$^2$,
\ba\la{eq:Rs}
&& R(s) = 
\theta(\sqrt{s}-2m_{\pi^\pm}) \; \theta(4.4m_{\pi^\pm}-\sqrt{s})
\\ && \qquad {\txts\frac{1}{4}}  
\Big[1-{\txts\frac{4m^2_{\pi^\pm}}{s}}\Big]^{3/2} 
\big(0.6473+ f_0(\sqrt{s})\big)
\nn && 
+ \theta(\sqrt{s}-4.4m_{\pi^\pm}) \theta(M_3 - \sqrt{s}) 
\left({\txts\sum_{i=1}^2} f_i(\sqrt{s})\right)
\nn &&
+ f_3(\sqrt{s}) + 3 \big( {\txts(\frac{2}{3})^2 + (\frac{1}{3})^2 
          +(\frac{1}{3})^2} \big) \theta(\sqrt{s}-M_3) .
\nonumber
\ea
To the $f_i$ correspond the parameters $\{C_i,M_i,\Gamma_i\}$
listed in \tab(\ref{tab:param}).

\begin{figure}
  \centering
  \includegraphics[width=0.45\textwidth]{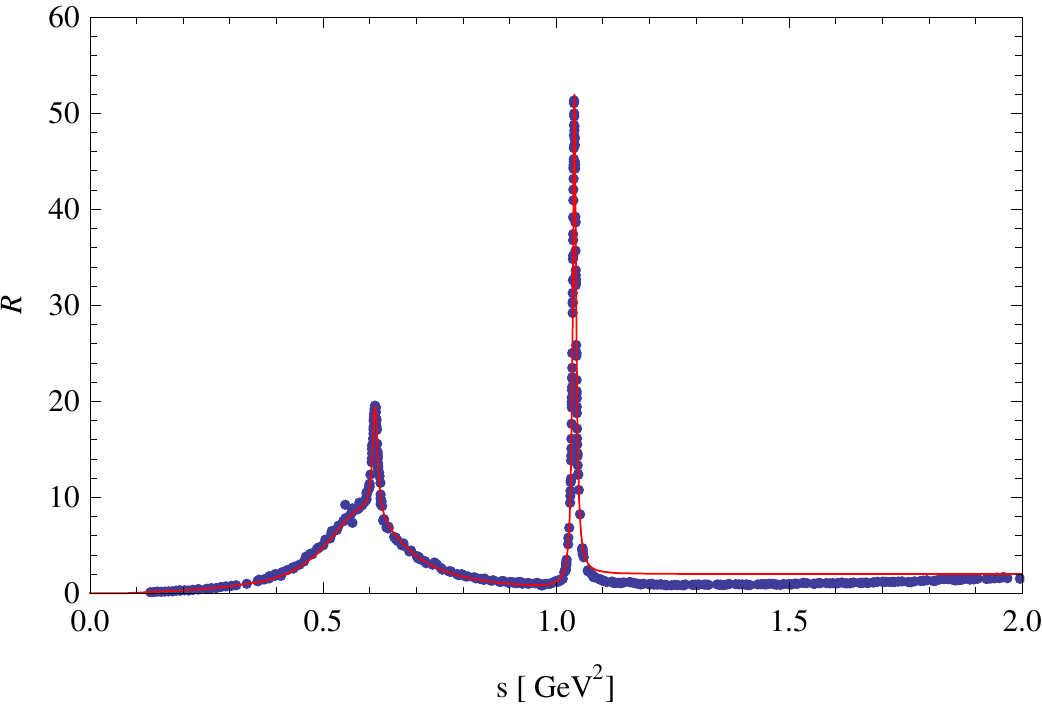}
  \caption{Parametrization of the $R(s)$ ratio.}
  \label{fig:RVergleich}
\end{figure}
\begin{figure}
  \centering
  \includegraphics[width=.45\textwidth]{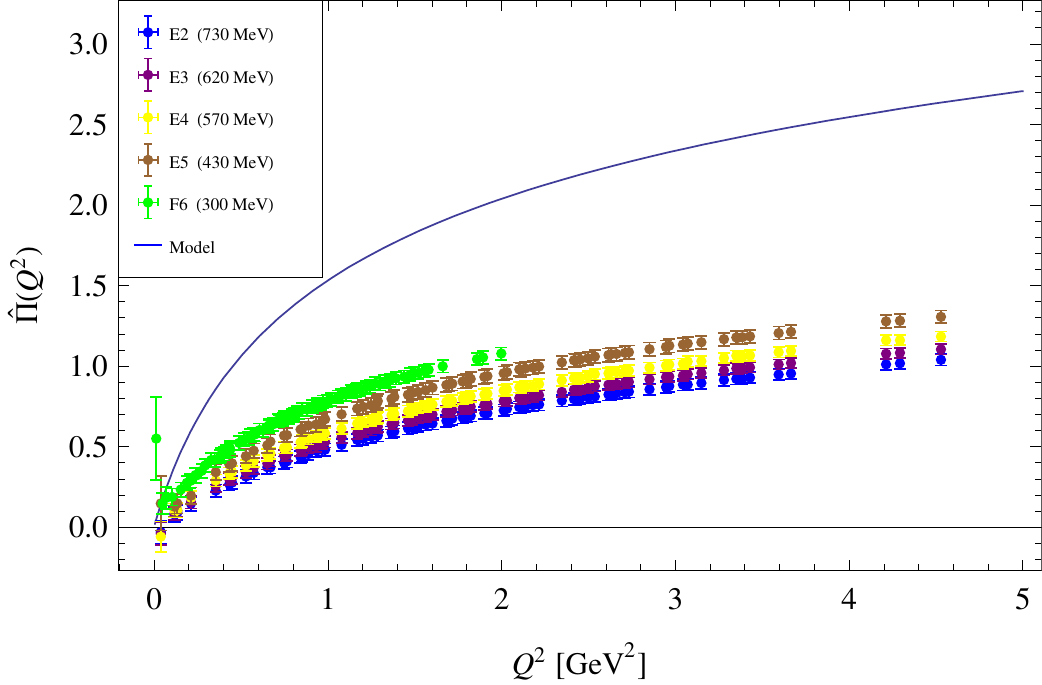}
  \caption{Comparison of the vacuum polarization calculated using our parametrization
(\ref{eq:Rs}) of the $R(s)$ ratio  and lattice data from~\cite{Brandt:2010ed,DellaMorte:2010sw}.}
  \label{fig:PiCompar}
\end{figure}

A comparison between the parametrization and the experimental data compiled by the 
PDG can be seen in figure \ref{fig:RVergleich}.
The vacuum polarization resulting from integrating our parametrization
of $R(s)$ in \eq(\ref{eq:DispRelR}) is shown in figure (\ref{fig:PiCompar}).
With the calculated vacuum polarization it is also possible to
calculate the hadronic contribution to the anomalous magnetic moment
of the muon $\amuHLO$ by using formula (\ref{eq:amublum2}).  Using
our simple parametrization of the $R(s)$ ratio, we obtaine the value
$\amuHLO=672\cdot10^{-10}$.  
It is close enough to the reference value of
$\amuHLO=(690.3\pm5.3)\cdot10^{-10}$~\cite{Jegerlehner:2008zz}
(see the review~\cite{Jegerlehner:2009ry} and the 
recent~\cite{Davier:2010nc,Hagiwara:2011af} for the latest evaluations) for
the purpose of this article to justify the simple functional form we
have used.

The phenomenological hadronic vacuum polarization is compared to
lattice QCD data generated by Della Morte et
al.~\cite{Brandt:2010ed,DellaMorte:2010sw}\footnote{A more extensive set of these
  lattice results was reviewed by J.~Zanotti in a plenary session of
  the Hadron 2011 conference, Munich, 13-17 June 2011.} in
\fig(\ref{fig:PiCompar}). The data is generated from $N_{\rm f}=2$
gauge field configurations, i.e.  they only include the sea quark
effects of the $u,d$ quarks.  In the electromagnetic current, the
contributions of the up, down and strange quarks were taken into
account. Results at different light-quark masses are displayed in
\fig(\ref{fig:PiCompar}), corresponding to `pion' masses $m_\pi$ down to
300MeV.  The E lattices are $32^3\times 64$ and the F lattices are
$48^3\times96$, so that the smallest value of $m_\pi L$ is 4.7. Only
the Wick-connected contributions were included in the calculation. Twisted
boundary conditions were used to obtain the vacuum polarization at a
denser set of momenta. For more details, we refer the reader to the
original publications~\cite{Brandt:2010ed,DellaMorte:2010sw}.

The subtracted vacuum polarization $\Pihat(Q^2)$ calculated on the
lattice lies about a factor two below the curve obtained using the
experimental $R(s)$ ratio. As the quark mass is lowered, the curves
(very slowly) approach the phenomenological curve. So in spite of the
excellent statistical quality of the lattice data, the large volumes
used, and the dense set of momenta, a large difference remains between
the lattice and the phenomenological curve. And this occurs in a
region (say, $0.4{\rm GeV}^2< Q^2 < 2.0{\rm GeV}^2$) where we would
expect cutoff effects to be small. We note however that in view of the
form of kernel $K(s)$, this region makes only a modest contribution to
$\amuHLO$, and that the smallness of $\amuHLO$ calculated around
$m_\pi=300$MeV is instead related to the behavior of $\Pi(Q^2)$ very
close to the origin (barely visible in \fig\ref{fig:PiCompar}).  In
fact, in the limit where the lepton mass goes to zero, the hadronic
contribution to its anomalous magnetic moment is given by
\be
\lim_{m_\mu\to0} \frac{\amuHLO}{m_\mu^2} = \frac{1}{3}
\left(\frac{\alpha}{\pi}\right)^2 \Big(\frac{\Pihat(s)}{s}\Big)_{s=0}.
\ee

Feng et al.~\cite{Feng:2011zk} followed an approach based on the idea
that using the variable $Q^2/m_\rho^2$ for the horizontal axis, where
$m_\rho$ is the quark-mass dependent $\rho$-meson mass, leads to an
approximate scaling at small $Q^2$, in the sense that the curves
corresponding to different quark masses would approximately lie on top
of eachother. This idea may prove helpful in carrying out the chiral
extrapolation, but it only partly explains the difference between 
the curves in \fig(\ref{fig:PiCompar}).

The different channels contributing to $R(s)$ and their relative
importance for the evaluation of the vacuum polarization and $\amuHLO$
have been described in the literature,
Refs.~\cite{Davier:2002dy,Davier:2010nc,Hagiwara:2011af,Jegerlehner:2009ry}
among others. For the reader's convenience we summarize some of the
facts known from $e^+e^-$ annihilation experiments that are relevant
to our discussion.  The total two-pion contribution represents
roughly $75\%$ of $\amuHLO$, and it is itself dominated by the $\rho(770)$
meson contribution: the part of it that comes from below 0.5GeV
amounts to 
$8.4\%$~\cite{AmelinoCamelia:2010me}, while the two-pion contribution
to $\amuHLO$ from $0.1\leq s\leq 0.85{\rm GeV^2}$ is
$69\%$~\cite{Muller:2009pj,Ambrosino:2010bv}.  
See in particular the accurate initial-state radiation (ISR) measurements
in this region~\cite{:2009fg,Ambrosino:2010bv}; 
the world's data is nicely summarized in Fig. (3) 
of Ref.~\cite{Hagiwara:2011af}.  The point has also been
made~\cite{Passera:2008jk} that if the energy region below 500MeV was
alone responsible for the current discrepancy between the Standard
Model prediction and the direct measurement of $\amuHLO$, it would
require a $52\%$ increase of the $e^+e^-\to\pi^+\pi^-$ cross section,
and it would then also lead to a lowered Higgs mass upper-bound of
143GeV at $95\%$ confidence level. Such an increase seems very
unlikely in view of the quoted experimental errors.

For comparison, the entire three-pion contribution to $\amuHLO$ from
threshold to 1.8GeV is about 
$6.8\%$~\cite{Hagiwara:2011af,Davier:2010nc} (see Fig.~(13)
of~\cite{Hagiwara:2011af}). From threshold to 660MeV, Hagiwara et
al.~\cite{Hagiwara:2011af} estimate it in Chiral Perturbation Theory
to be negligible, $0.001\%$.
The four-pion contribution has also been determined.
From 0.305 to 1.8GeV, the $2\pi^+2\pi^-$ channel yields $2.0\%$, 
while $\pi^+\pi^-2\pi^0$ yields $2.7\%$~\cite{Hagiwara:2011af}. 

We note that a lattice calculation containing just the Wick-connected
diagrams of the up and down quarks amounts to working with the isospin
current. Such a calculation can perfectly well be compared to the
experimental data if one selects unit isospin final hadronic
states. In particular, the $\omega$ and $\phi$ resonances should not
be included in the dispersion integral in such a comparison.  See
also~\cite{DellaMorte:2010aq} for an analysis of the Wick-disconnected
diagrams in chiral perturbation theory.

%% file: finivol.tex
\section{The vacuum polarization on the torus}
\la{sec:finivol}

As in all numerical lattice QCD calculations, the vacuum polarization is
evaluated in finite volume. We will assume that the boundary
conditions are periodic in all directions, except that the fermions
have antiperiodic boundary conditions in the time direction (the
system is thus at a finite, albeit low temperature). The question then
arises, how large the finite-size effects are, and what their
parametric dependence is.  A first step towards answering this
question is to provide a general parametrization of the polarization tensor
on the torus of dimensions $\beta\times L^3$.

The correlation function of a (not necessarily conserved) current
forms a rank-two symmetric tensor which can be uniquely decomposed
into a traceless and a scalar component,
\ba \la{eq:Pi}
\Pi_{\mu\nu}(q) &=& \bar \Pi_{\mu\nu}(q)  + \hat \Pi_{\mu\nu}(q),
\\
\hat \Pi_{\mu\nu}(q) &=& \delta_{\mu\nu} \hat\Pi(q),\qquad 
{\txts\sum_{\mu=0}^3} \bar \Pi_{\mu\mu}(q) = 0.
\ea
In infinite space ($\mathbb{R}^4$), the O(4) symmetry of the theory implies that 
\ba
\bar \Pi_{\mu\nu}(q)  &= & (q_\mu q_\nu -\frac{1}{d}\delta_{\mu\nu}q^2) f(q^2),
\\
\hat \Pi_{\mu\nu}(q) &=& \delta_{\mu\nu} q^2 g(q^2).
\ea
Imposing the conservation equation $q_\mu \Pi_{\mu\nu}(q)=0$, 
we obtain the condition 
\be
g(q^2) = -\frac{d-1}{d} f(q^2).
\ee
Returning to \eq(\ref{eq:Pi}), one then arrives at  
(\ref{eq:PimunuQ}),
\be \la{eq:Pimunu}
\Pi_{\mu\nu}(q) = (q_\mu q_\nu -\delta_{\mu\nu}q^2) f(q^2),
\ee
and $f(q^2)$ can be identified with the vacuum polarization, usually 
notated $\Pi(q^2)$.

On a four-dimensional torus of dimensions $L^4$,
the relevant symmetry group is the hypercubic group H(4).
We will follow the notation of Ref.~\cite{Gockeler:1996mu}.
The 20 inequivalent irreducible representations of H(4)
are denoted by $\tau_k^{(l)}$, where $l$ is the dimension of the representation
and $k=1,2,\dots$ distinguishes inequivalent representations of the same dimension.
There are four one-dimensional representations, two of dimension two; four of
dimension three, four, and six; and two of dimension eight.
The defining representation is labeled as $\tau^{(4)}_1$.
The currents $j_\mu(x)$ belong to this representation.
Their direct product can be decomposed according to~\cite{Gockeler:1996mu}
\be \la{eq:tau1xtau1}
\tau^{(4)}_1 \otimes \tau^{(4)}_1 = \tau_1^{(1)} \oplus \tau_1^{(3)} 
                                \oplus \tau_1^{(6)} \oplus \tau_3^{(6)}.
\ee
Because the vector correlator is symmetric in the spacetime indices $\mu$ and $\nu$,
the antisymmetric representation $\tau_1^{(6)}$ will play no role in the following.

Very often the space on which QCD is simulated is 
a four-dimensional torus of dimensions $\beta\times L^3$.
Then the symmetry group H(4) is further reduced to $Z(2)\times {\rm H}(3)$, 
where $Z(2)$ corresponds to Euclidean-time reversal and ${\rm H}(3)$ is the 
symmetry group of the cube. Two irreducible representations of H(4) appearing 
in \eq(\ref{eq:tau1xtau1}) further break up into smaller irreducible representations
of H(3). The latter are denoted $A_1$, $A_2$, $E$, $T_1$ and $T_2$ (respectively 
of dimensions 1, 1, 2, 3 and 3). We then have the decompositions
\ba
\tau_3^{(6)} &=& T_1 \oplus T_2,
\\
\tau_1^{(3)}  &=& A_1 \oplus E.
\ea
Thus 
\ba
\bar\Pi_{\mu\nu}(q) &=& \bar\Pi^{A_1}_{\mu\nu}(q) + \bar\Pi^{E}_{\mu\nu}(q)
+\bar\Pi^{T_1}_{\mu\nu}(q)+\bar\Pi^{T_2}_{\mu\nu}(q),
\\
\hat \Pi_{\mu\nu}(q)  &=& \delta_{\mu\nu} \hat\Pi^{A_1}(q).
\ea
In matrix notation, we now have (all matrices are symmetric)
\ba
\bar\Pi^{A_1}_{\mu\nu}(q)&=&  \left(\begin{array}{c@{~}c@{~}c@{~}c}
\bar\Pi^{A_1}_{00} & 0 & 0 & 0 \\
    &  \frac{-1}{3}\bar\Pi^{A_1}_{00} & 0  & 0\\
     &     &    \frac{-1}{3}\bar\Pi^{A_1}_{00} & 0 \\
 &  &  &  \frac{-1}{3}\bar\Pi^{A_1}_{00}
\end{array}  \right),
\\
\bar\Pi^{E}_{\mu\nu}(q) &=& 
\left(\begin{array}{c@{~~}c@{~~}c@{~~}c}
0  & 0  & 0 & 0 \\
  & \bar \Pi^{E}_{11} & 0 & 0 \\
&  &  \bar \Pi^{E}_{22} & 0 \\
 &    &    &   \bar \Pi^{E}_{33}
\end{array}  \right),
~ {\txts\sum_{j=1}^3} \bar\Pi^E_{jj}=0,
\\
\bar\Pi^{T_1}_{\mu\nu}(q) &=&  \left(\begin{array}{c@{~~}c@{~~}c@{~~}c}
0 & \bar\Pi^{T_1}_{01}  & \bar\Pi^{T_1}_{02} & \bar\Pi^{T_1}_{03} \\
  &  0  &  0  & 0 \\
  &     &  0  &  0 \\
&  &  &  0 
\end{array}  \right),
\\
\bar\Pi^{T_2}_{\mu\nu}(q)&=& \left(\begin{array}{c@{~~}c@{~~}c@{~~}c}
0 & 0  & 0 & 0 \\
  & 0  &  \bar\Pi^{T_2}_{12} & \bar\Pi^{T_2}_{13} \\
  &    &  0  &  \bar\Pi^{T_2}_{23} \\
  &    &     &   0    
\end{array}  \right).
\ea
Without using the conservation of the current, $\Pi_{\mu\nu}(q)$ is
thus characterized by five independent functions belonging to various
irreducible representations of the group H(3).

The Ward identities still read $q_\mu \Pi_{\mu\nu}(q)=0$ on the torus.
Writing them out for $\nu=0$ and 1,
\ba \la{eq:nu.eq.0}
q_0 \big(\bar\Pi^{A_1}_{00} + \hat\Pi^{A_1}\big) &=& 
-{\txts\sum_{j=1}^3} q_j \bar\Pi^{T_1}_{0j}  
\\ \la{eq:nu.eq.1}
q_1 \big( \hat\Pi^{A_1} + \bar\Pi^E_{11}- {\txts\frac{1}{3}} \bar\Pi_{00}^{A_1}\big)
\! &=&\!   -q_0 \bar\Pi^{T_1}_{01}
- q_2 \bar\Pi^{T_2}_{12} - q_3 \bar\Pi^{T_2}_{13}. ~~~~~
\ea
There are two further equations obtained from \eq(\ref{eq:nu.eq.1})
by cyclic permutation of the indices (1,2,3). The three representations
parametrizing the diagonal components can be related to the two off-diagonal
representations $T_1$ and $T_2$. For a generic momentum with 
non-vanishing components in each direction, we have the relations
\ba
4\bar\Pi_{00}^{A_1}(q) &=& 
\sum_{i\neq j} \bar\Pi^{T_2}_{ij} \,\frac{q_i}{q_j}
- \sum_j \bar\Pi_{0j}^{T_1} \big(3\frac{q_j}{q_0}-\frac{q_0}{q_j}\big),~~~
\\
4 \hat\Pi^{A_1}(q) &=& 
-\sum_{i\neq j} \bar\Pi^{T_2}_{ij} \, \frac{q_i}{q_j}
- \sum_j \bar\Pi^{T_1}_{0j}\big(\frac{q_j}{q_0}+\frac{q_0}{q_j}\big).~~~~~
\ea
If one or more components of the momentum vanish, one has to inspect 
the Ward identities on a case-by-case basis.
The component in the representation $E$ is obtained by taking the linear combination 
$q_2\cdot$\eq(\ref{eq:nu.eq.1}) $-q_1$\eq(\ref{eq:nu.eq.1})$_{q_1\to q_2\to q_3}$. 
The result is 
\ba
&&3\Pi^E_{11}(q)  = -{2} \frac{q_0}{q_1}\bar\Pi^{T_1}_{01} 
+ \frac{q_0}{q_2}\bar\Pi^{T_1}_{02} + \frac{q_0}{q_3}\bar\Pi^{T_1}_{03} 
\\ && 
+ \bar\Pi^{T_2}_{12} \Big(\frac{q_1}{q_2}-2\frac{q_2}{q_1}\Big)
+ \bar\Pi^{T_2}_{13} \Big(\frac{q_1}{q_3}-2\frac{q_3}{q_1}\Big)
+ \bar\Pi^{T_2}_{23} \Big(\frac{q_3}{q_2} + \frac{q_2}{q_3}\Big).
\nonumber
\ea
We note that for many actions at finite lattice spacing, the Ward identities read 
$\hat q_\mu \Pi_{\mu\nu}=0$ with $\hat q_\mu\equiv 2\sin(aq_\mu/2)$.
In equations (\ref{eq:nu.eq.0}) and (\ref{eq:nu.eq.1}), and in the equations in
the next subsection, $q_\mu$ should then be replaced by $\hat q_\mu$.

\subsection{Parametrization of the polarization tensor at small momenta}

Of particular relevance is the behavior of the polarization tensor at
very small momenta, since in infinite volume $\Pi(0)$ must be
subtracted from $\Pi(Q^2)$ in order to remove a logarithmic
divergence.  On the torus we Taylor-expand the different contributions to the
polarization tensor up to quartic order, including all the polynomials 
compatible with the respective cubic representations,
\ba \la{eq:poly1}
\bar\Pi^{T_1}_{01}(q) &=& 
q_0q_1 \big[ A_{T_1}+B_{T_1}\big(q_1^2-{\txts\frac{3}{5}}\bq^2\big)  
\\
&& \qquad + C_{T_1} q_0^2 + D_{T_1} \bq^2 \big],
\nn \la{eq:poly2}
\bar\Pi^{T_2}_{12}(q) &=& q_1q_2 \big[ A_{T_2}+ B_{T_2}q_0^2 + C_{T_2}\bq^2  
\\
&& \qquad +D_{T_2} \big( q_1^2+q_2^2 - {\txts\frac{6}{7}}\bq^2 \big)\big],
\nn \la{eq:poly3}
\bar\Pi^{E}_{11}(q) &=& A_E \big(q_1^2 - {\txts\frac{1}{3}}\bq^2\big) 
+ B_Eq_0^2 \big(q_1^2 - {\txts\frac{1}{3}}\bq^2\big) 
\\
&& + C_E \big(q_1^4 - {\txts\frac{1}{3}\sum_j} q_j^4 -{\txts\frac{6}{7}} \bq^2(q_1^2-{\txts\frac{1}{3}}\bq^2)\big)
\nn
&& +D_E \bq^2 \big(q_1^2 - {\txts\frac{1}{3}}\bq^2\big),
\nn \la{eq:poly4}
\bar\Pi^{A_1}_{11}(q) &=&  \bar Aq_0^2 + \bar B\bq^2 + \bar C q_0^4 + \bar Dq_0^2 \bq^2 
\\
&& + \bar E(\bq^2)^2 + \bar F \big({\txts\sum_j} q_j^4-{\txts\frac{3}{5}}(\bq^2)^2\big),
\nn \la{eq:poly5}
\hat\Pi^{A_1}(q) &=&  \hat Aq_0^2 + \hat B\bq^2 + \hat C q_0^4 + \hat Dq_0^2 \bq^2 
\\
&& + \hat E(\bq^2)^2 + \hat F \big({\txts\sum_j} q_j^4-{\txts\frac{3}{5}}(\bq^2)^2\big).
\nonumber
\ea
It should be noted that since the momentum variable $q$ assumes only
discrete values on the torus, one can always represent a function of
$q$ as a polynomial.  Also, in infinite volume, the spectral
representation (\ref{eq:DispRel}) shows that for a spectral function
admitting a mass gap, all the derivatives of the vacuum polarization
exist and are finite at the origin, and that the vacuum polarization
can be represented as a polynomial in $Q^2$ locally around the origin.
Therefore one expects those coefficients in
\eq(\ref{eq:poly1}--\ref{eq:poly5}) that are commensurate with the
infinite-volume tensor structure (\ref{eq:PimunuQ}) to smoothly tend
to a Taylor coefficient of $\Pi(Q^2)$ in the infinite-volume limit,
whereas the others must vanish in the same limit.
Before considering this limit however, we work out the consequences 
of the conservation of the current on the coefficients.

From the Ward identity (\ref{eq:nu.eq.0}), we
can express all five series expansions in terms of two of them, by
relating the coefficients.  We choose $\bar\Pi^{T_1}$ and
$\bar\Pi^{T_2}$ to be independent series and find for the other
coefficients
\ba \la{eq:Ec1}
A_E &=& A_{T_2},
\\
B_E &=& B_{T_2} - B_{T_1},
\\
C_E &=& 2 D_{T_2},
\\
D_E &=& C_{T_2} - {\txts\frac{1}{7}} D_{T_2};~~~~~~~~~~~~~~~~~~~~~~
\ea
\ba
\bar A &=& -{\txts\frac{1}{4}}  A_{T_1},
\\
\bar B &=& {\txts\frac{1}{4}}(A_{T_1} - {\txts\frac{2}{3}} A_{T_2}),
\\
\bar C &=& -{\txts\frac{1}{4}} C_{T_1},
\\
\bar  D &=& {\txts\frac{1}{4}} (C_{T_1} +{\txts\frac{4}{15}} B_{T_1} - D_{T_1} - {\txts\frac{2}{3}} B_{T_2}),
\\
\bar E &=& {\txts\frac{1}{4}} ( D_{T_1} +{\txts\frac{4}{105}} D_{T_2} - {\txts\frac{2}{3}} C_{T_2}  ),
\\
\bar F &=& {\txts\frac{1}{4}} (B_{T_1} - {\txts\frac{1}{3}}  D_{T_2} ) ;
\ea
\ba
\hat A &=& -{\txts\frac{3}{4}} A_{T_1} ,
\\
\hat B &=& -{\txts\frac{1}{4}} (A_{T_1} + 2A_{T_2}),
\\
\hat C &=& -{\txts\frac{3}{4}} C_{T_1},
\\
\hat  D &=& -{\txts\frac{1}{4}} ( -{\txts\frac{4}{5}} B_{T_1} + 3D_{T_1} + C_{T_1} + 2 B_{T_2}) ,
\\
\hat E &=& -{\txts\frac{1}{4}} ( -{\txts\frac{4}{35}}  D_{T_2} +2 C_{T_2}  + D_{T_1} ),
\\
\hat F &=& -{\txts\frac{1}{4}} ( B_{T_1} + D_{T_2} ) .
\la{eq:Eclast}
\ea
This exercise explicitly illustrates that two of the five introduced
functions are sufficient to fully characterize the polarization
tensor.  In particular, the latter is known entirely once the
off-diagonal components (parametrized by $\bar\Pi^{T_1}$ and
$\bar\Pi^{T_2}$) are known.

In addition to the terms listed above, there are terms that are not
constrained by the Ward identities, for instance 
\be
\Pi_{00}(0,\bq) \equiv \chi(\bq),
\ee
which can be interpreted as a static susceptibility.  When the time
extent $\beta$ is infinite, we have $\chi(\bq)\sim\bq^2$ at small
momentum, due to the presence of a mass gap in QCD, and this
guarantees that the photon remains massless. When $\beta$ is finite
however, $\chi(\boldsymbol{0})$ is the thermal electric charge
susceptibility, which does not vanish.
Our parametrization so far does not allow for this effect, and 
we must thus add the terms
\ba
\bar\Pi^{A_1}_{11} \supset \bar \chi \delta_{q,0},
~~~~~
\hat\Pi_{11}^{A_1} \supset \hat \chi \delta_{q,0}
\ea
to the generic expressions
 (\ref{eq:poly1}--\ref{eq:poly5})\footnote{At 
finite $\bq$, the coefficients $\hat B$ and $\bar B$ already 
capture the effect of the susceptibility $\chi(\bq)$.},
where $\hat \chi$ and $\bar \chi$ are two new coefficients.


When some of the dimensions become infinite, non-analyticities in the
momenta can appear. A familiar case is the limit where the spatial
dimensions become infinite, but $\beta$ remains finite. Then at small
momenta, $\Pi_{00}(q)\sim \frac{\chi(\boldsymbol{0})
  D\bq^2}{|q_0|+D\bq^2}$ due to the diffusion pole ($D$ is the
electric charge diffusion constant, see~\cite{Meyer:2011gj} for a
review).

In infinite volume, the vacuum polarization $\Pi(Q^2)$ is uniquely
defined, due to the constrained Lorentz structure of the polarization
tensor. On the $\beta\times L^3$ torus however, we see that there are
two obvious definitions of $\Pi(0)$, one that one may extract from the
$T_1$ representation, and one from the $T_2$ representation.
Indeed, to quadratic order the various components read
\ba
\Pi_{01}(q) &=& A_{T_1} q_0q_1 ,
\\
\Pi_{12}(q) &=& A_{T_2} q_1q_2 ,
\\
\Pi_{11}(q) &=& (\bar\chi + \hat\chi) \delta_{q,0} 
+ A_{T_2} q_1^2- ( A_{T_1}q_0^2 +A_{T_2} \bq^2),~~~~
\la{eq:Pi11smallq}
\\
\Pi_{00}(q) &=&   (\hat\chi -3 \bar\chi)\delta_{q,0}  -A_{T_1}\bq^2 .
\ea
Disregarding the susceptibility terms, they thus have the form
(\ref{eq:PimunuQ}) expected in infinite volume, except for $\Pi_{11}$
if $q_1^2\neq \bq^2$.  In general however $A_{T_1}\neq A_{T_2}$, and
one has to specify how the extrapolation to $q=0$ is done. The vacuum
polarization extracted from the component $\Pi_{00}(Q)$ is the same as
the one extracted from $\Pi_{01}$, but one must be careful with
$\Pi_{11}$, given that the tensor structure appearing in
\eq(\ref{eq:Pi11smallq}) is not the one expected in infinite volume
unless $A_{T_1}=A_{T_2}$.


The equalities (\ref{eq:Ec1}) to (\ref{eq:Eclast}) can be used to
test computer programs.  Once these tests are passed, a
number of checks for finite-volume effects can be made.  Apart
from $A_{T_1}\stackrel{?}{=}A_{T_2}$, one can check for instance
whether $B_{T_1}$ vanishes and $C_{T_1}\stackrel{?}{=}D_{T_1}$, as one
expects in infinite volume, and similarly in the $T_2$ channel.

Finally we note that if one calculates the isospin current correlator
and introduces twisted boundary conditions for the quarks fields (see
for instance~\cite{DellaMorte:2010sw} and Refs. therein), then the
cubic symmetry is in general broken down even further, and the
analysis above should be adapted accordingly.



\subsection{A sum rule for the finite-volume spectral function}

The Ward identities $k_\mu \Pi_{\mu\nu}=0$ imply 
\be
k_3 \Pi_{33}(k) = 0,\qquad k=(0,0,0,k_3).
\la{eq:WI33}
\ee
This relation holds both in finite and in infinite volume.  We now
want to turn it into a sum rule for the spectral function $\rho_{33}$
(introduced in \eq\ref{eq:rhoDef}) via the dispersion relation for the
Euclidean correlator. The spectral function associated with the
spatial current correlator $\Pi_{33}$ grows as $k_0^2$ at large
frequencies. Therefore a subtraction is necessary to obtain a
convergent sum rule. Taking the difference between the finite-volume
and the infinite volume correlator leads to a subtracted spectral
function $\propto k_0^{-2}$ at large frequencies, due to the absence of
operators of dimension less than 4 in the operator product expansion
of $\Pi_{\mu\nu}$. If $\Delta\rho_{33}$ is the difference between the
finite-volume spectral function and the infinite-volume one, the
spectral representation
\be
\Pi_{33}(k_3,\beta,L)
- 
\Pi_{33}(k_3,\infty,\infty)
 = \int_{-\infty}^\infty \frac{d\omega}{\omega} \Delta\rho_{33}(\omega,k_3,\beta,L) 
\ee
is thus convergent\footnote{We have written $\Pi$ and $\rho$ as functions of the 
momentum variables that do not vanish.}.
The Ward identity (\ref{eq:WI33}) can then be written
\be \la{eq:sr}
k_3 \int_{-\infty}^\infty \frac{d\omega}{\omega} 
\Delta\rho_{33}(\omega,k_3,\beta,L)  = 0.
\ee
For $k_3\neq 0$, this sum rule constrains the finite-volume
alterations of the spectral density relative to the infinite-volume
situation. It may be useful in the finite-temperature context, where
one wants to determine the distribution of the spectral weight
$\Delta\rho_{33}(\omega,k_3,\beta,L)$, see~\cite{Meyer:2011gj} and
references therein.  If one takes the spatial volume to infinity,
$k_3$ can be made as small as desired and the sum rule for
$\Delta\rho_{33}$ then also holds for $k_3=0$. We note that the sum
rule (\ref{eq:sr}) is simpler in QCD than in the ${\cal N}=4$ super
Yang-Mills theory, because in the latter case the contribution of the
scalar fields to the current contains a derivative, which means that
the current itself is not invariant under a local symmetry
transformation, and this leads to a contact term in the current
correlator~\cite{Baier:2009zy}.

%% file: piQ2.tex
\section{Computational strategies for $\Delta\alpha(Q^2)$ and $\amuHLO$}
\la{sec:mix}

In lattice QCD it is customary to work with correlation functions
which are functions of Euclidean time $t$ and spatial momentum $\bk$.
This representation has the advantage that the low-lying states
dominate exponentially at large $t$. Here we first consider the
situation in infinite volume, deriving the relation between the
Euclidean correlator and the spectral function $\rho(q^2)$ introduced
in \eq(\ref{eq:rhomunu}).  In the next section we show that $a_\mu$
can be calculated in terms of the $(t,\bk)$-dependent correlator using
a different kernel\footnote{H.M. is indebted to Andreas J\"uttner for
  bringing to his attention coordinate space methods for the
  calculation of $a_\mu$.}.  We propose a way of treating different
intervals of $t$ differently when computing $a_\mu$, combining the
Euclidean-space calculation with the low-energy part of the spectral
function.

\subsection{The pion form factor in the timelike region}

The pion form factor in the timelike region can in principle 
be calculated based on \eq(\ref{eq:result1}) in the threshold region,
\be \la{eq:valid}
2m_{\pi^\pm} \leq \sqrt{s} \leq 2\big(m_{\pi^\pm} + m_{\pi^0}\big).
\ee
Via \eq(\ref{eq:rhoR}), it can directly be compared to the
experimental results for the $R$ ratio (for a precision comparison,
care must be taken of QED corrections). The quality of the
experimental data can be viewed in Fig. (3) of
Ref.~\cite{Hagiwara:2011af}. A further constraint on this region comes
from the fact that the form factor $F_\pi(s)$ admits a convergent
Taylor expansion in $s$ with a radius given by the parameters of the $\rho$
meson, $(m_\rho,\Gamma_\rho)$. The first two terms are well constrained by the 
pion form factor in the spacelike region, 
\be
F_\pi(s) = 1 + {\txts\frac{1}{6}} r_\pi^2\cdot  s + {\rm O}(s^2).
\ee
This means that calculating the pion radius in lattice QCD also helps
to constrain/check the spectral function in the threshold region.
Although the region (\ref{eq:valid}) only accounts for about $10\%$ of
the total $\amuHLO$, it is still important to control its contribution
at the level of a few percent in preparation for the upcoming
$(g-2)_\mu$ experiment at Fermilab~\cite{E-989}.

This program would be a relatively modest but certainly valuable
contribution of lattice QCD to the precision determination of
$\Delta\alpha(M_Z^2)$ and $\amuHLO$.

\subsection{Mixed-representation correlator\la{sec:submix}}
We consider the positive-definite correlator\footnote{
In Minkowski space, the field operators 
$\hat j_\mu$ are hermitian, $\hat j_\mu^\dagger = j_\mu$.
In Euclidean space, while $\hat j_0^\dagger = \hat j_0$,
the spatial current is antihermitian $\hat j_z^\dagger = -\hat j_z$.}
\be \la{eq:GtDef}
G(t)\equiv  \int \ud\bx\; \<j^{\rm em}_z(t,\bx) j^{\rm em\,\dagger}_z(0)\>.
\ee
We will now derive a spectral representation for $G(t)$ in terms 
of the spectral function $\rho(q^2)$ of \eq(\ref{eq:rhomunu}).
To this end, we note that the correlator $G(t)$ can also be 
obtained from $\Pi_{\mu\nu}(q)$ by Fourier transformation,
\ba \la{eq:GtPikk}
G(t) &=& - \int_{-\infty}^\infty \frac{\ud\omega}{2\pi} 
\Pi_{zz}(\omega,\bk=0) \, e^{i\omega t}.
\ea
Now, the tensor structure 
(\ref{eq:PimunuQ}) implies 
\be \la{eq:PikkPi}
\Pi_{zz}(\omega,\bk=0)=-\omega^2\Pi(\omega^2),
\ee
and secondly we can substitute the dispersion relation (\ref{eq:DispRel})
into \eq(\ref{eq:GtPikk}).
Noting that the $\Pi(0)$ term only contributes for $t=0$, 
we obtain for $t\neq 0$
\be
G(t) = \int_{-\infty}^\infty \frac{\ud\omega}{2\pi} \omega^4 
\int_0^\infty\ud s \frac{\rho(s)}{s(s+\omega^2)}\, e^{i\omega t}.
\ee
The integral is easily carried out and one obtains (again for $t\neq 0$),
\ba
G(t) &=& \frac{1}{2} \int_0^\infty \ud s \;\sqrt{s} \rho(s) e^{-\sqrt{s}|t|}
\\ &=&
  \int_0^\infty \ud\omega \, \omega^2\rho(\omega^2) e^{-\omega |t|}.
\la{eq:GtRho}
\ea
\eq(\ref{eq:GtRho}) is the sought after spectral representation.
One can of course derive the spectral representation of the correlator 
with non-vanishing spatial momentum,
\ba \la{eq:jzjzq}
&& \int \ud\bx\, e^{-i\bk\cdot\bx}\, \<j_z(t,\bx)j_z^\dagger(0)\>
\\
&& \qquad = \int_{|\bk|}^\infty \ud\omega\, (\omega^2-\bk^2+k_z^2)
\, \rho(\omega^2-\bk^2) e^{-\omega |t|}.
\nonumber
\ea
Using the same parametrization of the phenomenological $R(s)$ ratio as
in section (\ref{sec:compar}), we obtain the dimensionless correlator
$t^3G(t)$ pictured in \fig(\ref{fig:t3Gt}). Beyond say 2fm, it falls
off rapidly to zero, asymptotically as $e^{-2m_\pi t}$. We note that
configuration-space Euclidean correlators have been used before to
confront instanton~\cite{Shuryak:1992ke} and holographic
models~\cite{Schafer:2007qy} with experimental data via dispersion
relations.

One can invert the Fourier transform (\ref{eq:GtPikk}), 
and express the vacuum polarization (\ref{eq:PikkPi})
through the mixed-representation correlator,
\be
\Pi(\omega^2) = \frac{1}{\omega^2}\int_{-\infty}^\infty \ud t\, e^{-i\omega t } G(t).
\ee
At small $\omega$, the vacuum polarization behaves as 
\be
\Pi(\omega^2) \stackrel{\omega\to0}{\sim } 
\frac{1}{\omega^2}\int_{-\infty}^\infty \ud t\,  G(t) - \frac{1}{2} \int_{-\infty}^\infty\ud t\, t^2 G(t)
+ \dots
\ee
The $\frac{1}{\omega^2}$ term vanishes, as the corresponding integral
represents the quark number susceptibility of the vacuum (if one
thinks of $z$ as the `time' direction).  Thus
\be
\Pi(\omega^2)-\Pi(0) = \int_{-\infty}^\infty\ud t\, G(t) 
\left[\frac{e^{-i\omega t}-1}{\omega^2} + \frac{t^2}{2}  \right].
\ee
This integral is UV-finite by power counting (leaving in the term 
$\int \ud t G(t)=0$ makes this explicit). We also note that 
$G(t)$ is an even function of $t$, so that the expression is real, 
and we can write
\ba
\Pi(\omega^2)-\Pi(0) &=& 
 2 \int_{0}^\infty\ud t\, G(t)
\Big[\frac{t^2}{2} - \frac{1-\cos{\omega t}}{\omega^2}  \Big]
\\  \la{eq:PihatGt}
&=&
\frac{1}{\omega^2} \int_{0}^\infty\!\!\!\!\ud t\, G(t)
\Big[\omega^2t^2 - 4\sin^2(\half\omega t) \Big].~~~~~
\ea

The leading hadronic contribution to the anomalous magnetic moment of
the muon can be expressed through the vacuum polarization $\Pi(Q^2)$, 
\eq(\ref{eq:amublum2} and \ref{eq:kerK}).
Using relation (\ref{eq:PihatGt}), we can write $\amuHLO$ as 
\ba  \la{eq:amu_t}
\amuHLO &=& {4\alpha^2\,m_\mu}\int_0^\infty \ud t \,
t^3 \;G(t) \; \tilde K(t),
\\
\tilde K(t) &\equiv& \frac{2}{m_\mu t^{3}} \int_0^\infty \frac{\ud\omega}{\omega} 
 \; K_E(\omega^2) \left[\omega^2 t^2  - 4\sin^2\big({\txts\frac{\omega t}{2}}\big)\right].
~~~~~
\ea
The kernel $\tilde K(t)$ is dimensionless, proportional to $t$ at
small $t$ and to $1/t$ at large $t$.  The factor $t^3$ was chosen
because $t^3 \;G(t)$ is dimensionless and finite as $t\to0$. The
integrand of \eq(\ref{eq:amu_t}) is displayed in
\fig(\ref{fig:t3GtKtilt}).

\subsection{Euclidean correlator vs. the $R$ ratio}

We have already confronted the vacuum polarization calculated on the
lattice to the $R(s)$ ratio via the dispersion relation in section
(\ref{sec:compar}). Here we propose to do the same in the mixed
$(t,\bk)$ representation. This has the advantage that the correlator
involves only on-shell states.  In addition, the continuum limit is
approached with O($a^2$) corrections (assuming the vector current is
improved).

\begin{figure}
\centering
\includegraphics[width=0.45\textwidth]{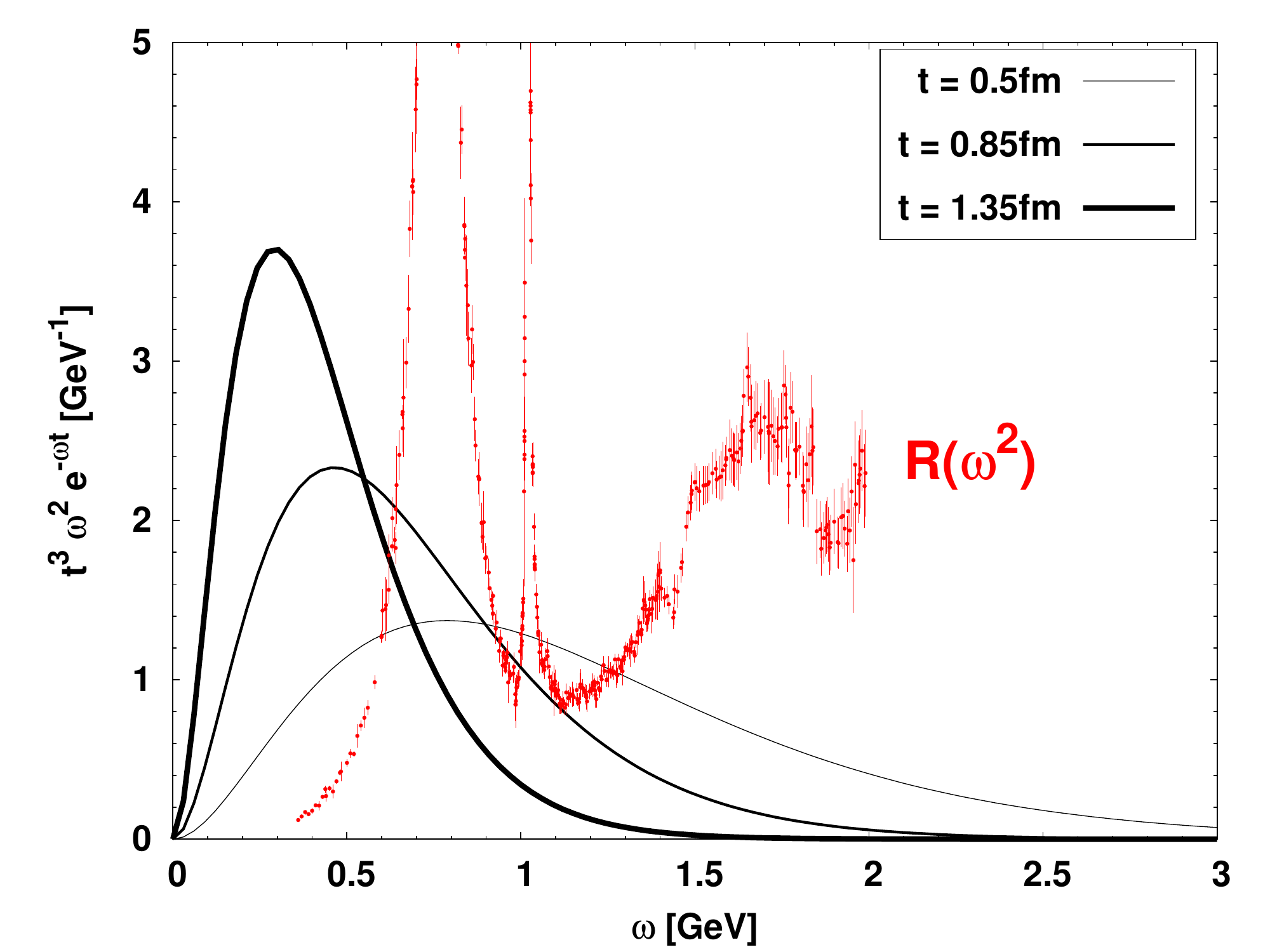}
\caption{The kernel of the spectral representation (\ref{eq:GtRho}) for 
three different Euclidean times, and the experimental $R(\omega^2)$ ratio 
up to $\omega=2$GeV~\cite{PDG2008}.}
\label{fig:t3w2exp}
\end{figure}

Consider the correlator $G(t)$, it is related to the spectral function
$\rho(\omega^2)$ via \eq(\ref{eq:GtRho}). Depending on the value of
$t$, the Euclidean correlator $G(t)$ is sensitive to different energy
intervals $\Delta\omega$. This simple fact is illustrated in
\fig(\ref{fig:t3w2exp}). Confronting in this way the value of $G(t)$
calculated on the lattice with the phenomenological value offers a way
for the lattice practitioner to test the validity of the $R(s)$
parametrization. Of course this test will only be useful if the
correlator, as well as its statistical and systematic error, are
accurately determined. This also means that the chiral extrapolation,
if any, must be under good control. Choosing $t$ between 0.5 and 1.5fm
yields sensitivity to the region most relevant to the muon anomalous
magnetic moment, encompassing in particular the $\rho$, $\omega$ and
$\phi$ resonances. The comparison can therefore provide a useful test
of the claimed accuracy of the phenomenological approach.

\subsection{Combining spacelike and timelike correlators} 

\begin{figure}
\centering
\includegraphics[width=0.45\textwidth]{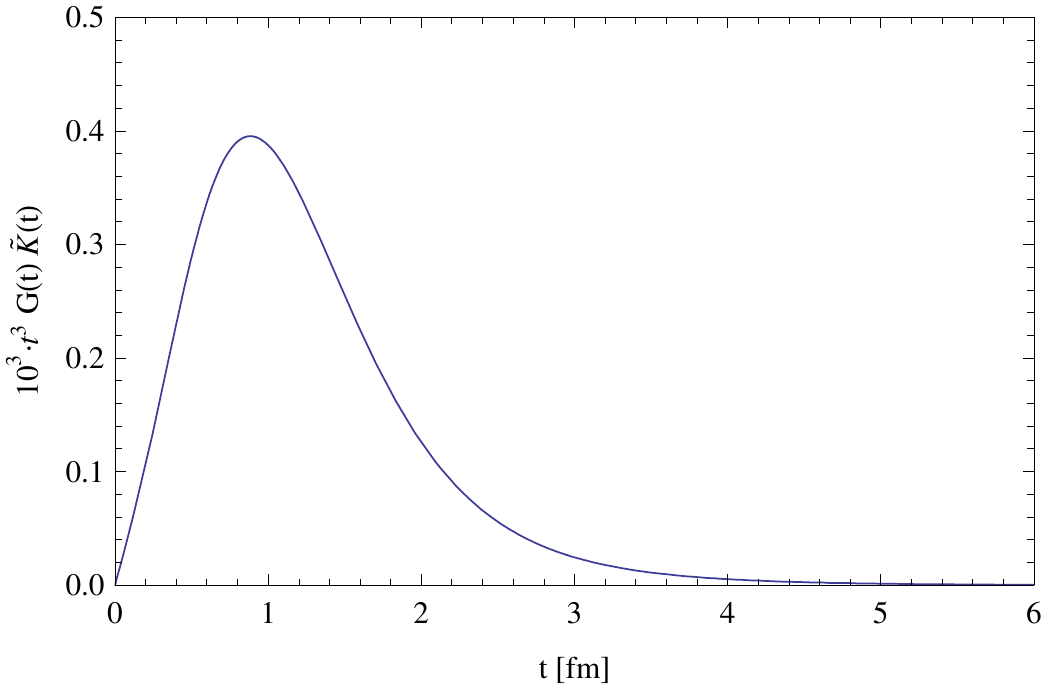}
\caption{The integrand of \eq(\ref{eq:amu_t}) in infinite volume, 
obtained from the phenomenological parametrization (\ref{eq:Rs}) 
of the $R(s)$ ratio.}
\label{fig:t3GtKtilt}
\end{figure}

There are many scales entering the integral (\ref{eq:amu_t}) yielding
$\amuHLO$.  The most relevant ones are the lepton mass $m_\mu$, the
pion mass $m_\pi$, the box size $L$ and the integration variable $t$.
An advantage of working in the mixed $(t,\bk)$ representation is that
a Hamiltonian interpretation of the expression is straightforward.
This motivates us to separate the contribution of different
$t$-intervals to the integral (\ref{eq:amu_t}) in the following
way. We write the integral as a sum of three terms,
\be
\amuHLO =  a^{<}_\mu(t_0,L) 
+ \Delta a_\mu(t_0,t_1,L) + a^{>}_\mu(t_1,L),
\ee
with 
\ba \la{eq:amu1}
a^{<}_\mu(t_0,L)&\equiv & {4\alpha^2\,m_\mu}\int_0^{t_0} \ud t \,
t^3 \;G(t) \; \tilde K(t) 
\\  \la{eq:amu2}
\Delta a_\mu(t_0,t_1,L)&\equiv & {4\alpha^2\,m_\mu}\int_{t_0}^{t_1} \ud t \,
t^3 \;G(t) \; \tilde K(t).
\\  \la{eq:amu3}
a^{>}_\mu(t_1,L)&\equiv & {4\alpha^2\,m_\mu}\int_{t_1}^\infty \ud t \,
t^3 \;G(t) \; \tilde K(t).
\ea
The short-distance contribution $a^<_\mu(t_0)\equiv\lim_{L\to\infty}
a^<_\mu(L,t)$ can be calculated in perturbation theory; the
perturbative series is known to have good convergence properties in
the vector channel. For the purpose of calculating $a_\mu$, the
practical question will be whether there is a choice of $t_{0}$ where
the perturbative series shows good convergence \emph{and} the
discretization errors are small\footnote{An analogous issue arises
  when computing $a_\mu$ in four-momentum space, where perturbation
  theory is found to match the lattice vacuum polarization down to
  about 3GeV$^2$~\cite{Aubin:2006xv}.}.

A state of energy $E$ makes a contribution of order $e^{-Et_1}$ to
$a^>_\mu(t_1,L)$.  With a solid understanding of the low-energy
spectrum, one can therefore analyze this contribution for large enough
$t_1$ and in particular its finite-volume effects. The larger $t_1$,
the stronger low-energy states dominate, but on the other hand the
numerical importance of this contribution is reduced. Therefore one
would like to choose $t_1$ as small as possible while the contribution
is still dominated by the analytically tractable low-lying states.

\begin{figure}[t]
\centering
\includegraphics[width=0.45\textwidth]{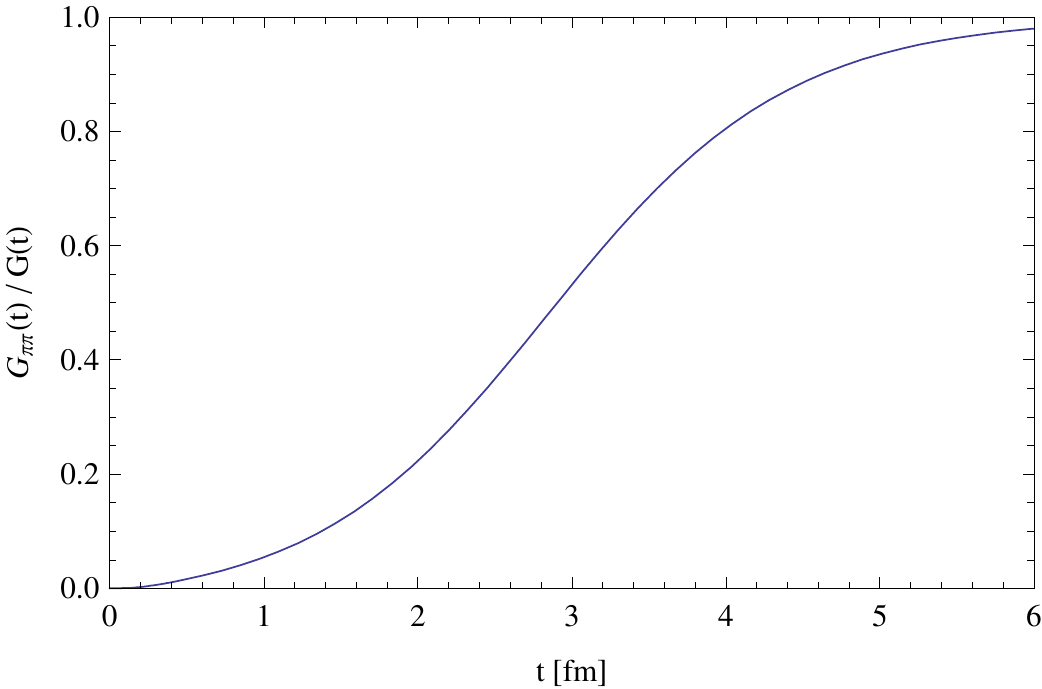}
\caption{The fraction of the $\bq=0$ Euclidean current correlator $G(t)$ coming from the region
$2m_{\pi^\pm}\leq\sqrt{s}<2(m_{\pi^\pm}+m_{\pi^0})$ of the spectral function
(in infinite spatial volume). See \eq(\ref{eq:GtRho}).}
\label{fig:GpipiG-HM}
\end{figure}

In appendix (\ref{sec:appFSE}) we show that the finite-size correction
on the contribution coming from $\rho(s<4(m_{\pi_\pm}+m_{\pi_0})^2)$ to the
long-distance part of the correlator is large (by long-distance, we
mean $t\gg (2m_\pi)^{-1}$ and $t\gg L/\pi$). Figure
(\ref{fig:GpipiG-HM}) illustrates that this contribution represents
$50\%$ of the full correlator starting at $t\approx 2.9$fm. The
contribution of $G(t\geq 2.9{\rm fm})$ to $\amuHLO$ is modest (see
\fig\ref{fig:t3GtKtilt}), but certainly not negligible if one aims at
a precision at the percent level or better on $\amuHLO$.  

In view of these finite-size corrections, it appears preferable to
treat the contribution $a^>_\mu(t_1)$ differently, if $t_1$ is chosen
large. \eq(\ref{eq:at0SR}) is an exact spectral representation of this
contribution. The infinite-volume spectral function $\rho(\omega^2)$
can be calculated on the lattice below the four-pion threshold via
\eq(\ref{eq:RFpi}, \ref{eq:result1}). Therefore, if the states below
the four-pion threshold saturate the current correlator beyond $t_1$,
$a^>_\mu(t_1)$ can be obtained from first principles through
\eq(\ref{eq:at0SR}), while $\Delta a_\mu$ is calculated by integrating
the Euclidean correlator as in (\ref{eq:amu2}). This is an example of
the potentially powerful interplay of a correlator in the spacelike
and in the timelike region. The difficulty with this strategy is that
numerically, the saturation at (say) the $95\%$ level by the states
below the four-pion threshold only begins around 5.3fm, at which point
the contribution to $\amuHLO$ is already extremely small. Choosing
$t_1$ as large as 5.3fm is therefore not a very attractive option.
The strategy of splitting up $\amuHLO$ into three contributions and
using the spectral representation for the long-distance part would
become more attractive if one could determine the spectral function up
to somewhat higher energies, but the formalism has not yet been developed
to handle the mixing of two-pion with four-pion states.

The analysis outlined above shows that the relative finite-size
effects on the long-distance contributions to the vacuum polarization
and to $\amuHLO$ are large. Fortunately, the size of these
contributions is small (albeit non-negligible) compared to the total
$\amuHLO$.  The finite-size effects associated with the higher-lying
states remains however unknown, and it should be investigated.

%% file: tau0.tex
\section{A new reference scale for lattice QCD}
\la{sec:tau0}

When making predictions for hadronic observables in lattice QCD, a
mandatory step is to calibrate the length of the lattice spacing $a$
in physical units (fm). In principle, calculating the proton mass
$M_p$ in lattice units, and equating it to 938.272MeV yields the
desired value. However, the proton mass is difficult to calculate
accurately on the lattice at light quark masses, and in practice one
chooses a different dimensionful quantity to `set the scale'. It is
however not easy to come up with a quantity which is both accurately
calculable on the lattice and accurately extracted from
experiment\footnote{If the goal is only to calibrate the
  \emph{relative} size of two lattice spacings, the second requirement
  is not mandatory.}. A quantity that has proved very useful is the
Sommer reference scale $r_0$, defined from the static quark
potential~\cite{Sommer:1993ce}. While it is accurately calculable on
the lattice, its value in the real world is not known precisely, since
its extraction from the upsilon spectrum requires introducing a
potential model. This procedure leads to a certain degree of ambiguity
(up to $\sim 8\%$, to be conservative).

\begin{figure}[t]
\centering
\includegraphics[width=0.45\textwidth]{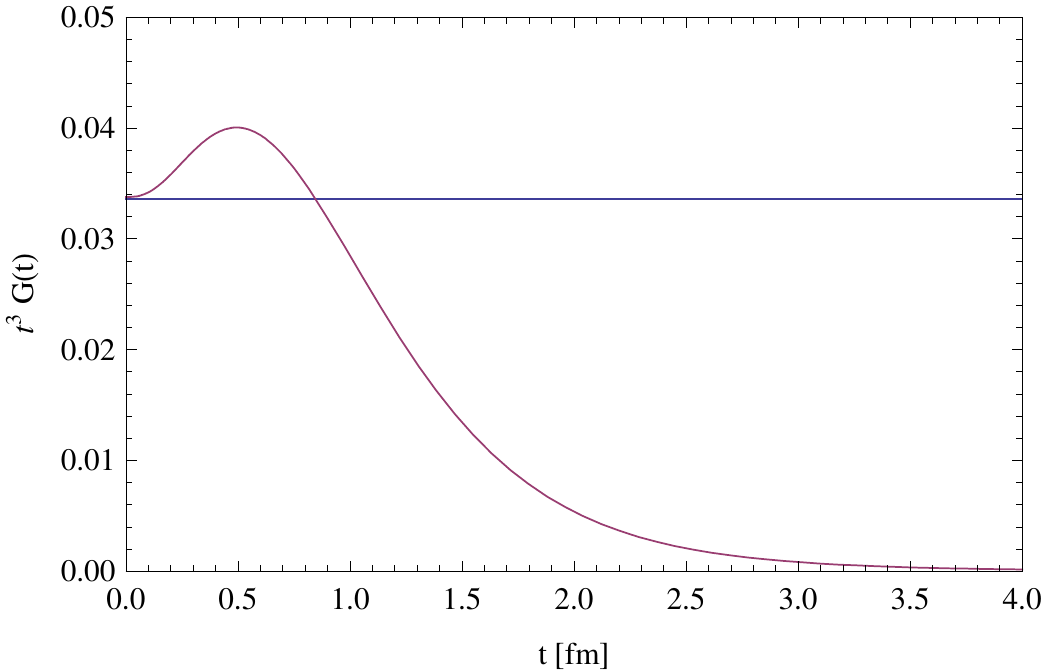}
\includegraphics[width=0.45\textwidth]{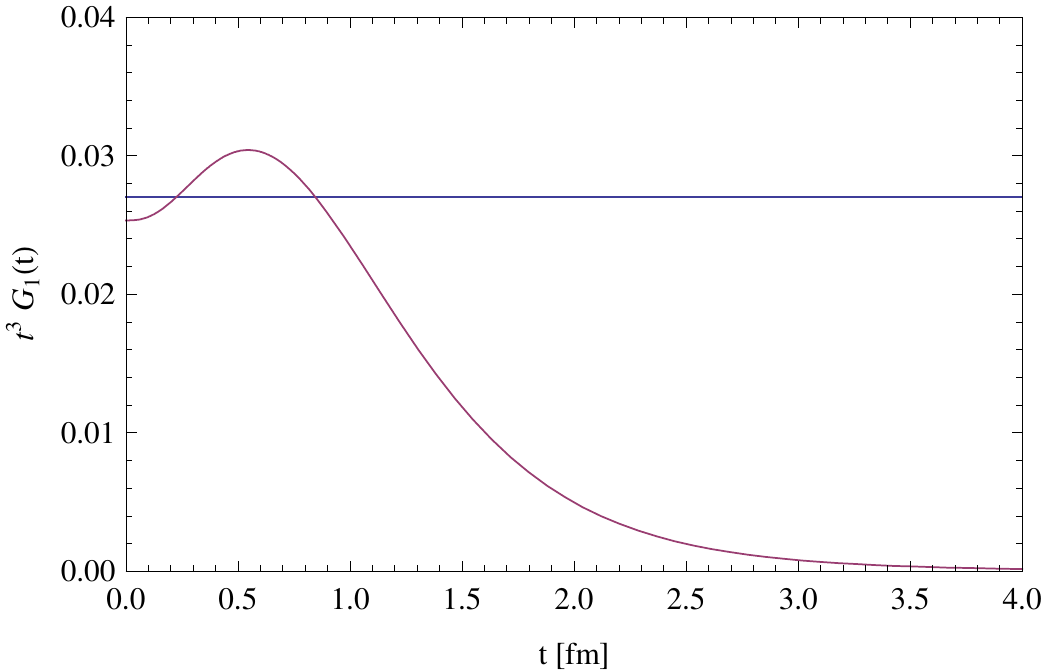}
\caption{Top: the correlator of the electromagnetic current $j_z^{\rm em}$ 
at zero spatial momentum in infinite volume obtained from the phenomenological 
parametrization of the $R(s)$ ratio. 
Bottom: the same for its isospin component $j_z^I$.}
\label{fig:t3Gt}
\end{figure}

Here we propose a new reference scale $\tau_0$ based on the vector
current correlator, which we believe satisfies the requirement of
being accurately calculable. Given the level of experimental effort
that has gone into the measurement of the $R(s)$ ratio particularly in
the past decade, we believe that the value of $\tau_0$ can also
accurately be extracted from the compiled data of $e^+e^-$
annihilation experiments.  Figure (\ref{fig:t3Gt}) displays the
electromagnetic current correlator $G(t)$ obtained from our simple
parametrization of the $R$ ratio, \eq(\ref{eq:Rs}).

At long time separations (somewhere beyond 1fm), the vector channel
correlator becomes noisy in Monte-Carlo simulations. At short
distances, one is confronted with cutoff effects from the lattice, and
secondly, QCD is approximately scale-invariant, so that the
sensitivity to the confinement scale is low. This dictates that one
should choose a reference time-scale somewhere between 0.5fm and 1fm.

Another computational aspect is that it is significantly easier to
calculate correlators in channels where no Wick-disconnected diagrams
appear. With the commonly used statistical sampling methods, connected
diagrams exhibit  better signal-to-noise ratios.  The
correlator of the electromagnetic current will contain the difference
of the strange-quark disconnected loops and the light-quark
disconnected loop.  Therefore, for the purpose of statistical accuracy
it is preferable to use the isospin current, in which disconnected
diagrams cancel out (assuming exact isospin symmetry). On the
phenomenological side, this means that one should only include those
final states with isospin $I=1$ in the evaluation of the $R$-ratio.
We therefore define 
\be
R_1(s) \equiv  \frac{\sigma(e^+e^-\to {\rm hadrons}\,|\; I=1)}
 {4\pi \alpha(s)^2 / (3s) } .
\ee
In particular, the bulk of the $\omega$ contribution, which
decays into an $I=0$ three-pion state, should not be included; nor
should the $\phi$ meson be included. The bulk of the low-energy
isospin-current spectral function will thus come from the two-pion
channel. The restriction to $I=1$ final states can be done in a
model-independent way as long as isospin breaking effects in QCD are
negligible to the desired degree of accuracy. The number of open
channels between 1.4GeV and 2.0GeV increases rapidly, and above 2GeV,
practically only inclusive measurements are made. Therefore it is
preferable to suppress the high-energy region by choosing a relatively
large $\tau_0$.

We thus define the Euclidean correlator $G_1(t)$ analogously to 
$G(t)$ (\eq\ref{eq:GtDef}), 
but replacing the electromagnetic current $j_\mu^{\rm em}$ 
by its isospin component
\be
j^I_\mu = \half (\bar u \gamma_\mu u - \bar d \gamma_\mu d).
\ee
We define the time scale $\tau_0$ by the equation
\be \la{eq:tau0}
\tau_0^3\, G_1(\tau_0) \stackrel{\rm def}{=} 0.027,
\ee
because it leads to $\tau_0$ in the desired range 0.5fm to 1.0fm. 

Figure (\ref{fig:t3Gt}) displays the electromagnetic current
correlator $G(t)$ obtained from our simple parametrization of the $R$
ratio, \eq(\ref{eq:Rs}), as well as the corresponding isospin current 
correlator $G_1(t)$ based on the model
\ba
 R_1(s) &=&
\theta(\sqrt{s}-2m_{\pi^\pm}) \; \theta(4.4m_{\pi^\pm}-\sqrt{s})
\\ && \qquad \times {\txts\frac{1}{4}}  
\Big[1-{\txts\frac{4m^2_{\pi^\pm}}{s}}\Big]^{3/2} 
\big(0.6473+ f_0(\sqrt{s})\big)
\nn && 
+ \theta(\sqrt{s}-4.4m_{\pi^\pm}) \theta(M_3 - \sqrt{s}) f_1(\sqrt{s})
\nn &&
 + 3 \big( ({\txts\frac{1}{2}})^2 + ({\txts\frac{1}{2}})^2 \big)\; \theta(\sqrt{s}-M_3) .
\nonumber
\ea
(the values used for the parameters are those in table \ref{tab:param}).
Based on this approximate form of the spectral function, we find 
\be
\tau_0 \simeq 0.85{\rm fm}.
\ee
We postpone a more accurate phenomenological evaluation of $\tau_0$
and its uncertainty to a future publication. At a practical level, a
nice feature of the $\tau_0$ definition is that no derivative must be
taken and no fit need be performed, one will `merely' have to perform an
interpolation to the point $\tau_0$.

Of course, many reference scales can be defined in a similar way.  
As a slight variation of the definition (\ref{eq:tau0}), one could also
use the time component of the current with a non-vanishing spatial
momentum $\bk$. The advantage is that the correlator does not vary as
fast at short distances, only as $\bk^2/t$ instead of $1/t^3$. The
spectral representation then reads\footnote{It is easily obtained 
from \eq(\ref{eq:jzjzq}) by using the Ward identity $\partial_\mu j_\mu=0$.}
\ba
&& \int \ud\bx\, e^{-i\bk\cdot\bx}\, \<j_0(t,\bx)j_0(0)\>
\\
&& \qquad = \bk^2 \int_{|\bk|}^\infty \ud\omega\,
\, \rho(\omega^2-\bk^2) e^{-\omega |t|}.
\nonumber
\ea
A reference scale based on this correlator might be an attractive 
alternative depending on its signal-to-noise ratio.

If one manages to calculate the disconnected diagrams accurately, then
a definition based on the electromagnetic current could become more
attractive.  For instance, defining $\tau_{\rm em}^3 G(\tau_{\rm em})
\stackrel{\rm def}{=} 0.0336$ leads again, using \eq(\ref{eq:Rs}), to
$\tau_{\rm em}\approx 0.85$. This has the advantage that the
selection of unit-isospin final states in the $R(s)$ ratio is not
required, nor is the assumption of exact isospin symmetry.

Finally, we remark that as with any another reference scale, a chiral
extrapolation is necessary unless a simulation is performed at
physical quark masses. We leave this question for future
investigation, but note that the properties of the $\rho$ meson
probably play an important role in this respect.

%% file: twisted.tex
\section{L\"uscher formula with twisted boundary conditions}
\label{sec:luescher3d}

For the evaluation of formula (\ref{eq:result1}) the scattering phase
shift $\delta_1$ of the $I=l=1$ channel must be known as a function to
obtain its derivative.  L\"uscher showed precisely~\cite{Luscher:1990ux} how
individual values of the scattering phase can be
reconstructed out of the two particle energy spectrum inside a finite
volume.  The problem that arises is that in practice, for a given volume size,
only a few values of the scattering phase shift can be reconstructed.
If the scattering phase shift is needed as a function of momentum the
computational cost increases rapidly since the results of many simulations with
different volume have to be combined.  Rummukainen and Gottlieb
\cite{Gottlieb} showed that by extending the formalism to non
vanishing center of mass momenta more points of the scattering phase
shift can be extracted per volume.  A further generalization is the use
of twisted boundary conditions that introduce a new continuous parameter, the
twist angle, to the simulations (see for instance~\cite{Luscher:1996sc}).  
The twist angle can then be used to modify the momentum almost continuously 
and therefore enables one to "scan" the scattering phase shifts.  
An elegant proof of the extended version of the L\"uscher formula 
was already given in~\cite{deDivitiis:2004rf}. Our derivation closely 
follows the original one in \cite{Luscher:1990ux} and is therefore only sketched here.  
First we introduce the wave function in infinite volume.  It describes
two bosons of equal mass, and obeys the Schr\"odinger equation
\begin{equation}
\bigg\lbrack\frac{\vec{p}^2}{2\mu}+V(\vert\vec{r}\vert)\bigg\rbrack\Psi(\vec{r})=E\,\Psi(\vec{r})
\end{equation}
where the potential $V(\vert\vec{r}\vert)$ is spherically symmetric and has a finite range $R$.
The wave function can be extended into spherical harmonics with its radial components satisfying the radial Schr\"odinger equation
\begin{equation}
    \bigg\lbrack\frac{d^2}{dr^2}+\frac{2}{r}\frac{d}{dr}-\frac{l(l+1)}{r^2}+k^2-2\mu V(r)\bigg\rbrack\Psi_{lm}(r)=0.
\end{equation}
In the region where the potential vanishes, the solution is given by 
$\Psi_{lm}(r)=\alpha_l(k)j_l(kr)+\beta_l(k)n_l(kr)$, where the constants $\alpha_l$
and $\beta_l$ determine the scattering phase shifts via
\begin{equation}
  e^{2i\delta_l(k)}=\frac{\alpha_l(k)+i\beta_l(k)}{\alpha_l(k)-i\beta_l(k)}.
\end{equation}
We now enclose both particles in a box of size $L\times L\times L$ and impose twisted boundary conditions
\begin{equation}
  \Psi(\vec{r}+\vec{n}L)=e^{i\vec{n}\cdot\vec{\phi}}\,\Psi(\vec{r}),\label{eq:tbc}
\end{equation}
where the triplet $\vec{\phi}$ of twist angles was introduced.
The potential must also be replaced by a periodic version 
\begin{equation}
V_L(\vec{r})=\sum_{\vec{n}\in\mathbb{Z}^3}V(\abs{\vec{r}+\vec{n}L}).
\end{equation}
Inside the box the energy spectrum is now discrete, 
with the energy-momentum relation still given by $E=k^2/2\mu$.
As in the periodic case, the Schr\"odinger equation 
in the region of vanishing potential $\Omega$ reduces 
to the Helmholtz equation,
\begin{equation}
  (\Delta+k^2)\Psi(\vec{r})=0.
\end{equation}
In this outer region $\Omega$ the eigenfunctions of the Hamiltonian
now have to be solutions of the Helmholtz equation, that are
expandable in spherical harmonics and have radial components
\begin{equation}
\Psi_{lm}(r)=b_{lm}\big(\alpha_l(k)j_l(kr)+\beta_l(k)n_l(kr)\big).
\end{equation}
According to a theorem by L\"uscher (proof given in appendix A of
\cite{Luscher:1990ux}) for each solution of the Helmholtz equation in
$\Omega$ that can be expanded in this way there exists a unique
eigenfunction of the Hamiltonian that matches this solution in
$\Omega$.  Finding the general solution of the Helmholtz equation
 therefore suffices for determining the formula for the scattering
phase shifts.

To simplify the derivation a bit, the momenta of the solutions of the
Helmholtz equation are now assumed not to belong to the singular set $
\Gamma_s=\big\lbrace
k\in\mathbb{R}\,\big\vert\,k=\pm\frac{2\pi}{L}\abs{\vec{n}},\,\text{for
}\vec{n}\in\mathbb{Z}^3\big\rbrace$.  Momenta in this set would allow
plane waves as solutions of the Helmholtz equation, which would
complicate the derivation a bit. For the following steps an angular
momentum cutoff $\Lambda$ is introduced, so that only the partial
waves with an angular momentum smaller or equal to $\Lambda$ feel the
presence of the potential.

The solutions we are about to construct should satisfy two conditions.
First they have to satisfy the twisted boundary conditions
(\ref{eq:tbc}), and second they should be bounded by a power of
$r^{-1}$ near the origin
\begin{equation}
  \lim_{r\rightarrow0}\,\abs{r^{\Lambda+1}\Psi(\vec{r})}<\infty.
\end{equation}
The ansatz for finding the general solution is now the Greens function
\begin{equation}
  \Gphi=L^{-3}\sum_{\vec{p}\in\Gamma^\phi}\frac{e^{i\,\vec{p}\cdot\vec{r}}}{p^2-k^2}
\end{equation}
where the momenta are elements of $
\Gamma^\phi=\big\lbrace\vec{p}\in\mathbb{R}^3\,\Big\vert\,
\vec{p}=\frac{2\pi}{L}\vec{n}+L^{-1}\vec{\phi},\;\vec{n}\in\mathbb{Z}^3\big\rbrace$.
That this function satisfies the twisted boundary conditions can be
shown straightforwardly, and it is also a solution of the Helmholtz
equation for $\vec{r}\neq0\,(\operatorname{mod}L)$.  By comparison
with the spherical Bessel functions one finds the behaviour of $\Gphi$
near the origin to be
\begin{equation}
  \Gphi=\frac{k}{4\pi}n_0(kr)+\hat G^\phi(\vec{r};k^2)
\end{equation}
where $\hat G^\phi(\vec{r};k^2)$ is the regular part.
Further solutions may be constructed by using the function 
$\mathcal{Y}_{lm}(\vec{x})=r^lY_{lm}(\theta,\phi)$ to obtain the derivatives
\begin{equation}
  \Glm=\mathcal{Y}_{lm}(\vec{\nabla})\Gphi.
\end{equation}
One can then show that the $\Glm$ form a complete, normal basis of the
solutions of the Helmholtz equation.  Therefore the solutions that
were searched for can be constructed as a linear combination of the
$\Glm$
\begin{equation}
  \Psi(\vec{r})=\sum_{l=0}^\Lambda\sum_{m=-l}^lv_{lm}\Glm.
\end{equation}
Like in L\"uscher's paper these solutions now have to be expanded in
spherical harmonics and this form has to be compared to the expansion
containing the spherical Bessel functions.  The expansion is similar
to the one with periodic boundary conditions
\begin{equation}
\begin{split}
  \Glm=&\frac{(-1)^lk^{l+1}}{4\pi}\Big\lbrace n_l(kr)Y_{lm}(\theta,\phi)\\ &+
\sum_{l'=0}^\infty\sum_{m'=l'}^{l'}\mathcal{M}_{lm,l'm'}^\phi\big(q(k)\big)j_{l'}(kr)Y_{l'm'}(\theta,\phi)\Big\rbrace,
\end{split}
\end{equation}
where $q=\frac{kL}{2\pi}$ and $\mathcal{M}_{lm,l'm'}^\phi\big(q(k)\big)$ is given by
\begin{equation}
  \mathcal{M}_{lm,l'm'}^\phi(q)=\frac{(-1)^l}{\pi^{3/2}}\sum_{j=\abs{l-l'}}^{l+l'}\sum_{s=-j}^j\frac{i^j}{q^{j+1}}C_{lm,js,l'm'}\mathcal{Z}_{js}^\phi(1;q^2).
\end{equation}
While the coefficients $C_{lm,js,l'm'}$ are given in
\cite{Luscher:1990ux} and can easily be calculated for a given set of
indices, the Zeta-functions $\mathcal{Z}_{lm}^\phi(1;q^2)$ (defined below) 
need to be determined numerically.  From the comparison of the expansions one
finds (for the choice $\boldsymbol{\phi}=(\phi,\phi,\phi)$ so as not to break the cubic symmetry) 
the usual L\"uscher formula
\begin{equation}
  \det\bigg\lbrack e^{2i\delta}-\frac{\mathbf{M}^\phi({\cal R})+i}{\mathbf{M}^\phi({\cal R})-i}\bigg\rbrack=0\label{eq:deltared}
\end{equation}
where now the $\mathbf{M}^\phi$ are determined by a different
Zeta-function.  This formula is already the reduced one that was
obtained by projecting on one of the irreducible representations
(${\cal R}=A_1^\pm,\,A_2^\pm,\,E^\pm,\,T_1^\pm,\,T_2^\pm$) of the
cubic group, as is described in \cite{Luscher:1990ux}.  The remaining
task now is to calculate the Zeta-functions
$\mathcal{Z}_{lm}^\phi(s;q^2)$ that are defined by
\begin{equation}
  \mathcal{Z}_{lm}^\phi(s;q^2)=\sum_{\vec{r}\in\tilde\Gamma^\phi}\mathcal{Y}_{lm}(\vec{r})(\vec{r}^2-q^2)^{-s}
\end{equation}
for $\Re{s}>1$, with $\tilde\Gamma^\phi=\big\lbrace\vec{r}\in\mathbb{R}^3\big\vert\vec{r}=\vec{n}+(2\pi)^{-1}\vec{\phi},\,\vec{n}\in\mathbb{Z}^3\big\rbrace$.
Like in L\"uschers original work the Zeta-functions are best calculated numerically using integral representations of the form 
\begin{equation}
\begin{split}
  \mathcal{Z}_{lm}^\phi(1;q^2)=
  &\sum_{\substack{\abs{\vec{r}}<\lambda\\ \vec{r}\in\tilde\Gamma^\phi}}\mathcal{Y}_{lm}(\vec{r})(\vec{r}^2-q^2)^{-1}\\ &+
(2\pi)^3\int_0^\infty\!dt\bigg\lbrack e^{tq^2}\mathcal{H}_{lm}^{\phi,\lambda}(t,0)-\frac{\delta_{l0}\delta_{m0}}{(4\pi)^2t^{3/2}}\bigg\rbrack.
\end{split}
\end{equation}
Here the reduced heat-kernel
$\mathcal{H}_{lm}^{\phi,\lambda}(t,\vec{x})$ of the Laplace operator
on a torus with twisted boundary conditions of size $L=2\pi$ appears,
and $\lambda$ must be chosen so that $\lambda^2>\Re q^2$.  Depending
on the value of $t$ the heat kernel has two different representations
based on
\begin{equation}
\begin{split}
  \mathcal{H}^\phi(t,\vec{x})&=(4\pi t)^{3/2}\sum_{\vec{n}\in\mathbb{Z}^3}e^{i\vec{n}\cdot\vec{\phi}}e^{-\frac{1}{4t}(\vec{x}-2\pi\vec{n})^2}\\
&=(2\pi)^{-3}\sum_{\vec{r}\in\tilde\Gamma^\phi}e^{i\vec{r}\cdot\vec{x}-t\vec{r}^2},
\end{split}
\end{equation}
the first converges for $t\leq1$ and the second one for $t\geq1$.
With those two representations the reduced version of the heat-kernel that is needed for the integral representation of the Zeta-function can then be defined by
\begin{equation}
\begin{split}
  \mathcal{H}_{lm}^{\phi,\lambda}(t,\vec{x})=(-i)^l\mathcal{Y}_{lm}(\vec{\nabla})\Big\lbrack&\mathcal{H}^\phi(t,\vec{x})\\ &-(2\pi)^{-3}\sum_{\substack{\abs{\vec{r}}<\lambda\\ \vec{r}\in\tilde\Gamma^\phi}}e^{i\vec{r}\cdot\vec{x}-t\vec{r}^2}\Big\rbrack.
\end{split}
\end{equation}
Using for instance the program {\it Mathematica} the Zeta-function can then be calculated numerically.
\begin{figure}
  \centering
  \subfloat[periodic boundary conditions]{\label{fig:Delta0nTW}\includegraphics[width=0.45\textwidth]{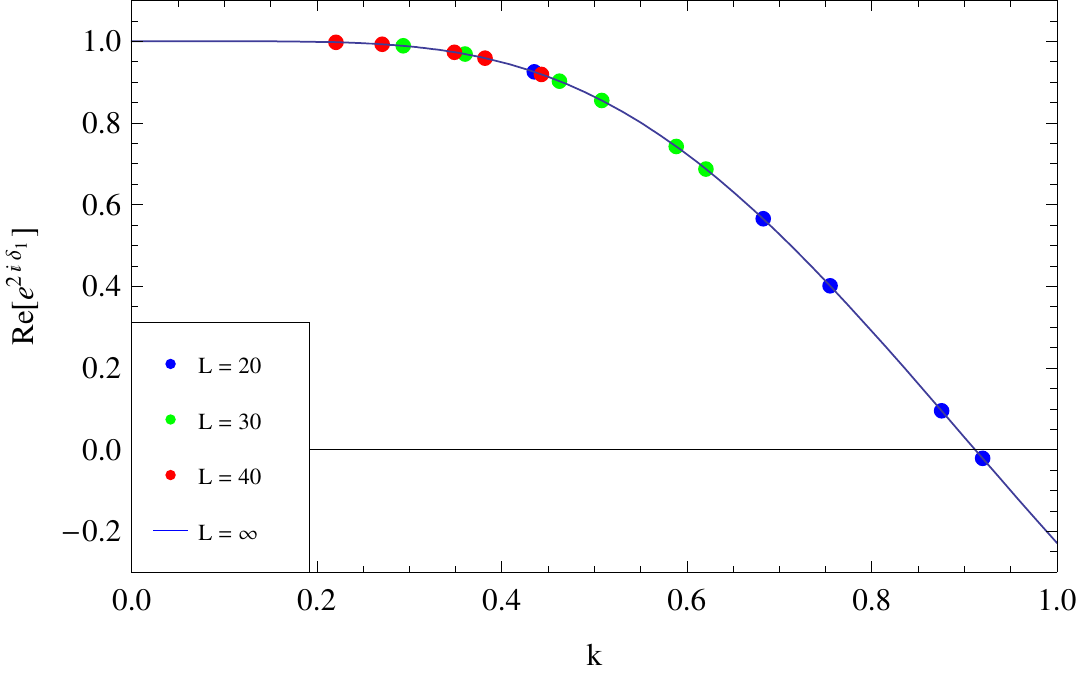}}\qquad\qquad
\subfloat[twisted boundary conditions ($L=20$)]{\label{fig:Delta0TW}\includegraphics[width=0.45\textwidth]{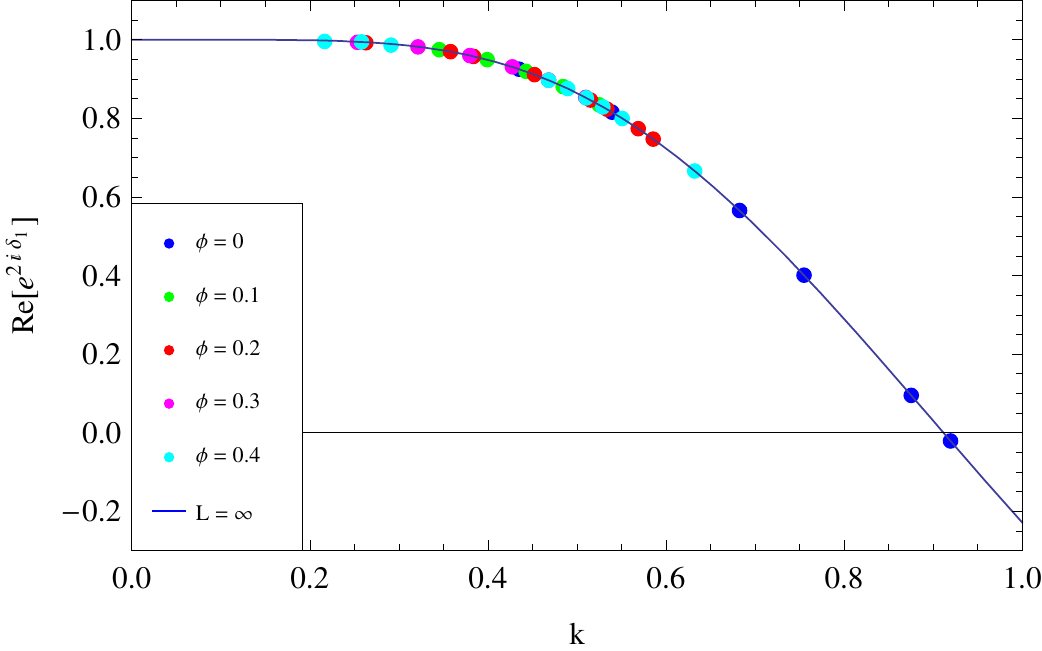}}
  \caption{The reconstructed scattering phase shift $\delta_1(k)$ using the L\"uscher formula with twisted boundary conditions. For futher information on this example see appendix \ref{app:beispiel}}
  \label{fig:nTW-TW}
\end{figure}

Figure \ref{fig:nTW-TW} illustrates the advantage of using the twisted
boundary conditions.  For a simple quantum mechanical example with a
potential well (see appendix \ref{app:beispiel}) the reconstructed
scattering phase shift $\delta_1(k)$ is
shown. Fig. \ref{fig:Delta0nTW} shows the scattering phase shift
reconstructed using periodic boundary conditions and several volumes,
while for fig. \ref{fig:Delta0TW} twisted boundary conditions were
used.  In the second figure the volume was set to be $L=20$, while the
twist angle ranges between $\phi=0$ and $\phi=0.4$.  In both figures
the blue curve is the scattering phase shift calculated in infinite
volume that was added as a comparison to the numerically calculated
data.  This simple example already shows the advantages of the twisted
boundary conditions, since the number of points that could be
reconstructed from one single volume could be improved.  Also the
distances between the individual points is much smaller, so that is
would also be possible to calculate the derivative of the scattering
phase shift, which is needed for equation (\ref{eq:result1}).

%% file: concl.tex
\section{Conclusion}

Given the highly accurate determinations of the $R$ ratio in $e^+e^-$
experiments, where now even the dominant isospin breaking effects are
taken into account, it is a challenge for lattice QCD to have a
phenomenological impact on the determination of the hadronic vacuum
polarization.  In view of the upcoming $(g-2)_\mu$ experiment at
Fermilab~\cite{E-989}, an accuracy of well below $1\%$ is called for on
$\amuHLO$. We have outlined two alternatives to the direct calculation
of the vacuum polarization. One option is the direct determination of
the spectral function $\rho(s)$ in the threshold region along the
lines of~\cite{Meyer:2011um}.  We described the determination of the
scattering phases in the vector channel using twisted boundary
conditions in section (\ref{sec:luescher3d}).  A second promising
approach is the calculation of the vector correlator $G(t)$ (defined
in \eq\ref{eq:GtDef}) in the region 0.5fm to 1.5fm or so, which
provides a check on the $R$ ratio in the energy region that makes the
largest contribution to $\amuHLO$.  We also considered the option of
integrating $G(t)$ from $t_0$ to $t_1$, with the short-distance
contribution treated in perturbation theory and the long-distance
contribution obtained from the pion timelike form factor via the
spectral representation, but it appears that the spectral function
would have to be determined up to higher energies than the four-pion
threshold for this strategy to be practical.

As a new idea we proposed to turn the table around and exploit the
accurate knowledge of the vector spectral function to define a
reference scale $\tau_0$ in QCD which can be determined accurately and
reliably. We believe that the only real difficulty could arise from
finite-size effects, which should therefore be investigated numerically
and analytically.

We have made first steps to study the finite-size effects affecting
the calculation of the vacuum polarization and the hadronic
contribution to $(g-2)_\mu$.  In section (\ref{sec:finivol}), we have
analyzed the tensor structure of the polarization tensor on the torus,
and find that for generic values of the momentum, two functions
invariant under the transformations of the cubic group $H(3)$
characterize the vector current correlator. This allows one to perform
various tests for finite-size effects. Based on current conservation,
we also derived a sum rule for the difference of the infinite-volume
spectral function and the finite-temperature/finite-volume
spectral function, \eq(\ref{eq:sr}). Then focussing on the spatial
current correlator projected onto zero spatial momentum, we analyzed
the contributions of different states as a function of Euclidean
time. At large times, the relative finite volume effects become order
unity, due to the discreteness of the two-pion states on the
torus. This contribution however only represents a few percent of
$\amuHLO$, due to the dominance of the $\rho$ resonance in this
quantity. We plan to study the finite-size effects of this
contribution in the near future.


%% file: appendixFSE.tex
\section{Finite-volume effects on the low-energy contribution to $\amuHLO$}
\la{sec:appFSE}

In dealing with the different scales of the problem we will consider that 
\be
\frac{m_\pi }{m_\mu} = {\rm O}(1).
\ee
Because of this relation, the kernel $\tilde K(t)$ cannot be expanded,
neither in a small-$t$ expansion, nor in a large-$t$ expansion.

In the chiral regime, the lowest-lying states contributing are
two-pion states. Let us call the ground state energy $E_0$. Roughly at
$E_0+2m_\pi$ begin the states with a four-pion component in their
wavefunction which is not exponentially small in the volume. Since it
is not known how to relate the relative weight of these components to
inelastic $\pi\pi$ scattering in infinite volume, we choose $t_1$
in such a way that the contribution of these states is suppressed,
\be
2m_\pi t_1 \gg 1.
\ee
We assume that the box is large, in the sense that 
\be
m_\pi L \gtrsim \pi.
\ee
If we also choose $t_1$ such that 
\be
t_1 \gg \frac{L}{\pi},
\ee
then only the ground state contributes to $G(t)$ for $t\geq t_1$.
Since all factors in the integrand are slowly varying, except for the
exponential $e^{-Et}$, the contribution to $a_\mu$ is approximately 
\be \la{eq:>L}
a^>_\mu(L,t_1) \simeq 4\alpha^2 m_\mu t_1^3 \tilde K(t_1) |A_0(L)|^2 \frac{e^{-E_0t_1} }{E_0},
\ee
where $A_0=L^{3/2}\<{\rm vac}|j(\bx)|\psi_0\>$ is the matrix element of the vector current 
between the lowest-lying (two-pion) state in the box and the vacuum; it is related
to the pion form factor through \eq(\ref{eq:result1}).
In infinite volume, the situation is different, because there are states 
arbitrarily close to $2m_\pi$, and we must integrate over them.
We define 
\be
a^>_\mu(t_1) = \lim_{L\to \infty}  a^>_\mu(L,t_1).
\ee
Using the spectral representation of the correlator, one finds
\ba\la{eq:at0SR}
a^>_\mu(t_1) 
&=& 4\alpha^2 m_\mu \int_0^\infty\!\!\! \ud \omega\; \omega^2 \rho(\omega^2)
\int_{t_1}^\infty\!\!\! \ud t \, t^3 \, \tilde K(t)\, e^{-\omega t}~~~~~~
\\ &\simeq  &
4\alpha^2 m_\mu \tilde K(t_1) t_1^3 
\int_0^\infty \!\!\!\ud \omega\; \omega \rho(\omega^2)  e^{-\omega t_1}.
\ea
The large value of $t_1$ dictates that only the small $\omega$ region
contributes, where the spectral density coincides with the timelike pion factor,
see \eq(\ref{eq:RFpi}). One then finds
\be 
\la{eq:>Infty}
a^>_\mu(t_1) \simeq  4\alpha^2 m_\mu t_1^3 \tilde K(t_1) 
\frac{|F_\pi(2m_\pi)|^2}{48\pi^2} \, 
\frac{3\sqrt{\pi}}{2} \frac{e^{-2m_\pi t_1}}{m_\pi^{1/2} t_1^{5/2}}.
\ee
Comparing \eq(\ref{eq:>L}) and (\ref{eq:>Infty}), we see that
the expression obtained in finite volume is parametrically different, because
$\exp(-E_0(L)t_1) \ll \exp(-2m_\pi t_1)$. The finite-volume effect is 
order unity for this long-distance contribution.

If the box is very large, $m_\pi L \gg \pi$, the contribution $\Delta
a_\mu(L,t_1)$ for $t_1\lesssim \frac{L}{\pi}$ (but still $2m_\pi
t_1\gg 1$) will be affected only by a small relative finite-volume
effect.  In this regime, the energy gaps between the two-pion states
is much smaller than $m_\pi$, so that the sum over states is a good
approximation to the integral over momenta in infinite volume. This
regime was investigated in~\cite{Lin:2001ek}. However, this regime is
hardly realistically achievable with present computing resources,
because the condition $m_\pi L \gg \pi$ implies that $L$ must be 10fm
at the very least when $m_\pi$ is set to its physical value.

%% file: potwell.tex
\section{Illustration with a potential well}
\label{app:beispiel}

In section (\ref{sec:luescher3d}) the scattering phase shifts calculated
for the quantum mechanical potential well were shown as an example.
This appendix describes how these scattering phase shifts were
calculated.

The potential well is of the form
\begin{equation}
  V(\abs{\vec{r}})=-\alpha\cdot\Theta(R-\abs{\vec{r}}),
\end{equation}
where $R$ is the range of the potential and $\alpha$ its strength.
For a comparison to the scattering phase shifts calculated using 
the L\"uscher formula the scattering phase shift in infinite volume is needed.
To solution to the Schr\"odinger equation
\begin{equation}
  \bigg\lbrack-\frac{1}{2\mu}\Delta+V(\abs{\vec{r}})\bigg\rbrack\Psi(\vec{r})=\frac{k^2}{2\mu}\Psi(\vec{r})
\end{equation}
is best defined piecewise
\begin{equation}
  \Psi(\vec{r})=\sum_{l=0}^\infty\sum_{m=-l}^lY_{lm}(\theta,\phi)
  \begin{cases}
    \alpha_lj_l(kr)+\beta_ln_l(kr)&r\geq R\\
    j_l(\tilde kr)&r\leq R    .
  \end{cases}
\end{equation}
Requiring smoothness of the solution and its derivative at $r=R$ 
leads to a system of equations for the coefficiants $\alpha_l(k)$, $\beta_l(k)$.
Once these are calculated the scattering phase shift can be calculated using the formula
\begin{equation}
  \delta_l(k)=\arctan\bigg\lbrack\frac{\beta_l}{\alpha_l}\bigg\rbrack.
\end{equation}

In order to calculate the scattering phase shifts 
using the formula derived in section (\ref{sec:luescher3d}) 
the first step now is to calculate the spectrum in the box.
The general solution of the Helmholtz equation 
is used to define a solution piecewise inside the box,
\begin{align}
  \Psi(\vec{r})\Big\vert_{\abs{\vec{r}}\geq R}
  &{}=\sum_{l=0}^\Lambda\sum_{m=-l}^lv_{lm}\Glm, \\
  \begin{split}
    \Psi(\vec{r})\Big\vert_{\abs{\vec{r}}\leq R}&{}
    =\sum_{l=0}^\Lambda\sum_{m=-l}^la_{lm}Y_{lm}(\theta,\phi)
  j_l(\tilde kr)\\
&{}\phantom{=}+\sum_{l=\Lambda+1}^\infty\sum_{m=-l}^la_{lm}Y_{lm}(\theta,\phi)
  j_l(kr).
  \end{split}
\end{align}
The advantage of this ansatz is that the general solution
 of the Helmholtz equation that is used already satisfies
 the twisted boundary conditions
\begin{equation}
  \Psi(\vec{r}+\vec{n}L)=e^{i\,\vec{n}\cdot\vec{\phi}}\Psi(\vec{r}),\;\;\vec{n}\in\mathbb{Z}^3.
\end{equation}
To simplify the calculation and to study the effects of higher 
scattering phase shifts the angular momentum cutoff $\Lambda$ 
was introduced in the same way as in section (\ref{sec:luescher3d}).
Before matching both pieces of the solution at the boundary $r=R$,
the solution outside the range of the potential can be rewritten as
\begin{align}
  \Psi(\vec{r})&{}=\sum_{l=0}^\Lambda\sum_{m=-l}^lv_{lm}\Glm&\\
  \begin{split}
    &{}=\sum_{l'=0}^\infty\sum_{m'=-l'}^{l'}Y_{l'm'}(\theta\phi)
    \Big\lbrace \\ & ~~~ j_{l'}(kr)\sum_{l=0}^\Lambda\sum_{m=-l}^{l}v_{lm}
    \frac{(-)^lk^{l+1}}{4\pi}\mathcal{M}^\phi_{lm;l'm'}\\
    &{}\phantom{=}+\Theta(\Lambda-l')v_{l'm'}
    \frac{(-)^{l'}k^{l'+1}}{4\pi}n_{l'}(kr)\Big\rbrace.
  \end{split}
\end{align}
Since the $Y_{lm}(\theta,\phi)$ are normal to each other one gets a
homogeneous system of equations when equating the two pieces of the
solution at $r=R$.  The important point to note is that the equations
for $l>\Lambda$ do not influence the spectrum, they merely determine
the $a_{l>\Lambda,m}$ in terms of the $v_{l\leq\Lambda,m}$. The latter coefficients 
are determined by the equations $l\leq\Lambda$. One has $\Lambda+1$
homogeneous equations for as many variables.  According to a well
known theorem a non-trivial solution to a set of homogeneous equations
only exists when the determinant of the corresponding matrix 
vanishes. This condition yields the allowed $k$-values.  For
$\Lambda=2$ for instance, the determinant of the matrix of
coefficients is of the block-diagonal form
\begin{equation}
  \begin{vmatrix}
    \mathbf{\Lambda0}&0&0\\
    0&\mathbf{\Lambda1}&0\\
    0&0&\mathbf{\Lambda2}
  \end{vmatrix}=\det\lbrack\mathbf{\Lambda0}\rbrack\cdot\det\lbrack\mathbf{\Lambda1}\rbrack\cdot\det\lbrack\mathbf{\Lambda2}\rbrack=0
\end{equation}
so that it seperates.
Since we are interested in the scattering phase shift 
$\delta_1$ we only need to consider the determinant 
$\det\lbrack\mathbf{\Lambda1}\rbrack$.
The system of equations that needs to be solved is
\begin{align}
   0&=v_{1-1}\frac{k^2}{4\pi}\big\lbrack j_1(kR)\mathcal{M}_{1-1;1-1}+n_1(kR)\big\rbrack+a_{1-1}j_1(\tilde kR)\\
  0&=v_{10}\frac{k^2}{4\pi}\big\lbrack j_1(kR)\mathcal{M}_{10;10}+n_1(kR)\big\rbrack+a_{10}j_1(\tilde kR)\\
  0&=v_{11}\frac{k^2}{4\pi}\big\lbrack j_1(kR)\mathcal{M}_{11;11}+n_1(kR)\big\rbrack+a_{11}j_1(\tilde kR).
 \end{align}
In this example the roots of the determinant were found by plotting
the determinant as a function of $k$ using {\it Mathematica}
and then using the function \verb+FindRoot+.  The calculated $k$ values 
can then be inserted into the L\"uscher formula for twisted boundary
conditions derived in section (\ref{sec:luescher3d}) to calculate the
scattering phase shifts.  The parameters that were used to obtain the
figures \ref{fig:Delta0nTW}, \ref{fig:Delta0TW} were $\alpha=2$, $R=2$
and $\mu=1$.